\documentstyle[verditz,epsf,rotate]{lamuphys}
\makeatletter
\let\chapter\hid@chapter
\makeatother
\begin{document}



\title{
{\it\small
LA-UR 98-4082 to appear in the Proceedings of 
the 1st International Conference on Quasiclassical Methods in 
Superconductivity, eds. D. Rainer and J.A. Sauls, 
Verditz, Austria (1998) \\[4ex]
}
{Electronic Transport in Unconventional Superconductors}
}

\author{Matthias\,J.\,Graf}
\institute{Center for Materials Science,
	Los Alamos National Laboratory,\\
  	Los Alamos, New Mexico 87545, USA}

\authorrunning{Matthias\,J.\,Graf}
\titlerunning{Electronic Transport in Unconventional Superconductors}

\maketitle

\begin{abstract}
We investigate the electronic transport coefficients
in unconventional superconductors at low temperatures,
where charge and heat transport are dominated by electron scattering
from random lattice defects. 
We discuss the features of the pairing symmetry, Fermi surface, and
excitation spectrum which are reflected in the low temperature heat
transport. For temperatures $k_B T \la \gamma \ll \Delta_0$, 
where $\gamma$ is the bandwidth of impurity induced Andreev
states, certain eigenvalues become {\it universal}, i.e.,
independent of the impurity concentration and phase shift.
Deep in the superconducting phase ($k_B T \la \gamma$)
the Wiedemann-Franz law, with Sommerfeld's value of the Lorenz
number, is recovered. We compare our results for theoretical models of
unconventional superconductivity in high-T$_c$ and heavy fermion
superconductors with experiment.  Our findings show that impurities
are a sensitive probe of the low-energy excitation spectrum, and
that the zero-temperature limit of the transport coefficients
provides an important test of 
the order parameter symmetry.\footnote{This work was
done in collaboration with S.-K.~Yip, J.~A.~Sauls, and D.~Rainer.}
\end{abstract}

\section{Introduction}

Heat and charge transport in metals provides
valuable information on the spectrum of charge
carriers and phonons, as well as the the scattering of
the carriers by defects and impurities.
In a normal metal at low temperatures the contribution of
the phonons to the electrical and thermal conductivities becomes
negligible compared to the electronic part.  When the temperature is 
further reduced the electronic transport is dominated by 
scattering of electrons from lattice defects.  Eventually the
electrical conductivity approaches a constant, $\sigma(T) \propto
\mbox{\sl constant}$, while the thermal conductivity becomes linear in
temperature, $\kappa(T) \propto T$.  Both transport coefficients are 
related by the Wiedemann-Franz (WF) law
$\kappa/(\sigma\,T) = \Ls$, with Sommerfeld's result for the Lorenz number
$\Ls = \frac{\pi^2}{3} (k_B/e)^2$.  Inelastic scattering leads to 
violations of the WF law and to significant deviations from $\Ls$.

  This picture is dramatically altered upon entering the superconducting
phase with the formation of a pair condensate and the opening of
a gap in the excitation spectrum. The excitation gap in conventional 
($s$-wave) superconductors destroys the WF law.
This effect has been known 
since the early days of the BCS (Bardeen, Cooper \& Schrieffer 1957) 
theory of superconductivity from the work of 
Mattis and Bardeen (1958) on the electrical conductivity, and from
the analysis of Geilikman (1958) of the heat transport.
  
  In this article we investigate the behavior of the heat current for an
unconventional superconductor, \ie, for an order parameter with reduced
symmetry for which gapless excitations exist even at zero temperature.
Such superconducting states have been argued to exist both in the
cuprates and heavy fermion superconductors.  
[For reviews on the heavy fermion system \UPt\ see, e.g., Sauls (1994a), 
and Heffner \& Norman (1996), and on the cuprates, e.g., Pines (1994a), and
Scalapino (1995)].
The leading pairing candidate for a tetragonal crystal structure (D$_{4h}$) 
in the cuprates is the B$_{1g}$ state, a spin-singlet state with lines of nodes
at the Fermi positions $p_{\!fx} = \pm p_{\!fy}$.  The most
promising candidates in the heavy fermion system UPt$_3$ are 
the two-dimensional orbital representations coupled to a symmetry 
breaking field.
For UPt$_3$, which has a hexagonal crystal structure ($D_{6h}$), phase 
diagram studies (Bruls \et\ 1990, Adenwalla \et\ 1990),
and transport measurements (Shivaram \et\ 1986, M\"uller \et\ 1987,
Broholm \et\ 1990, Signore \et\ 1995) lead to either an
even-parity, spin singlet E$_{1g}$ state, or an odd-parity, spin triplet
E$_{2u}$ pairing state. 
  In both cases the order parameter vanishes at
the Fermi surface on a line in the basal plane, $p_{\!fz}=0$, and at 
points at the poles, $p_{\!fx}=p_{\!fy}=0$.
Other models studied in the Ginzburg-Landau regime include the
one-dimensional orbital, spin triplet model with weak spin-orbit coupling
by Machida \& Ozaki (1991) and Ohmi \& Machida (1993), the accidental
degeneracy AB-model with two unrelated one-dimensional orbital
representations by Chen \& Garg (1993, 1994), and the accidental
degeneracy AE-model by Zhitomirsky \& Ueda (1996).

It is known that in a {\it clean} superconductor with an order 
parameter that vanishes along a line on the Fermi surface the low-energy
density of states is linear in the excitation energy,
$N(\epsilon)\sim N_{\!f}\,\epsilon/\Delta_0$ for
$\epsilon<\Delta_0$, which leads at low temperatures to a power law 
behavior in transport properties as a function of temperature for
$T\ll T_c$ (Coffey \et\ 1985, Pethick 
\& Pines 1986, Barash \et\ 1996).
Gor'kov \& Kalugin (1985)
and later Choi \& Muzikar (1988)
showed that this spectrum is altered by a random distribution of 
impurities.  A new low energy scale, $\gamma$, emerges
below which the density of states is approximately constant
and non-zero at zero energy. The energy scale $\gamma$ is interpreted
as the bandwidth of quasiparticle states bound to
impurities (Shiba 1968, Rusinov 1969, Buchholtz \& Zwicknagl 1981, Preosti
\et\ 1994, Balatsky \et\ 1995).
These impurity-induced states develop below the superconducting
transition. They are formed by the constructive interference
of particle- and hole-like excitations that undergo Andreev scattering
from the anisotropy of the order parameter in momentum space.
For an order parameter with a line of nodes that is required by
symmetry, the bandwidth and density of Andreev states at zero energy are
finite for any finite concentration of impurities,
$n_{i}\ne 0$. Both $\gamma$ and $N(0)$
depend on the impurity concentration $n_{i}$ and the
scattering phase shift $\delta_0$. Thus, $\gamma$ defines a
crossover energy scale or crossover temperature $T^\star$, 
below which the transport properties
are dominated by the Andreev
states. For excitation energies above $\gamma$ the transport
properties are determined primarily by the scattering of continuum
excitations.

  We study in detail the low temperature behavior of the thermal conductivity 
tensor for unconventional superconductors with line and point nodes in 
the order parameter.  One of the issues we address is the universality 
of the electrical and thermal conductivities at low temperature.  
We show that the components of the thermal conductivity tensor
that depend on quasiparticles near the line nodes
are determined by the same scattering rate as the
electrical conductivity and are universal in the limit $T\to 0$.
Furthermore, we show that the WF law is recovered
in the limit $T \ll T^\star$,
independent of the universality of the transport coefficients.
However, a significant temperature dependence of the Lorenz ratio
occurs over the temperature range, $T^\star \la T < T_c$, even for 
elastic scattering. The universal eigenvalues for $\tensor{\kappa}$ and 
$\tensor{\sigma}$ result from the cancellation between two factors:
(i) the density of Andreev states, which is proportional to
$N(\epsilon\!\to\! 0) \sim N_{\!f}\,\gamma/\Delta_0$, and (ii) the
reduction of phase space for scattering of gapless excitations, which
is proportional to
$\tau(\epsilon\!\to\! 0) \sim \hbar/\gamma$.  This leads
to estimates for the electrical conductivity,
$\sigma(T) \sim N(\epsilon)\,v_{\!f}^2\, \tau(\epsilon)
\sim N_{\!f} v_{\!f}^2 (\hbar/\Delta_0)$, and the thermal conductivity,
$\kappa(T) \sim N(\epsilon) k_B^2\,T\,v_{\!f}^2\, 
\tau(\epsilon) \sim N_{\!f} v_{\!f}^2 k_B^2\,T (\hbar/\Delta_0)$,
which are independent of the defect density or scattering phase shift.
Perhaps the most surprising result is that the ratio of the
thermal and electrical conductivity gives the Sommerfeld
value for the Lorenz ratio, $\kappa/\sigma T \simeq L_{\!S} =
\frac{\pi^2}{3}(k_B/e)^2$.  Thus, the differences in the coherence
factors that determine the conductivity tensors, $\tensor{\kappa}$
and $\tensor{\sigma}$, do {\it not} affect the Lorenz ratio
$L(T)=\kappa/\sigma T$ for $T\ll T^\star$ and $\hbar\omega\ll\gamma$.
This discovery was made independently by Sun \& Maki (1995) for a
two-dimensional $d_{x^2-y^2}$ superconductor.
For temperatures above the crossover region, $T > T^\star$, the
Lorenz ratio deviates significantly from the Sommerfeld value.  
The WF law breaks down due to the energy dependence of the
scattering lifetime in the superconducting state, although scattering 
of electrons from impurities is purely elastic. 

The rest of this article is organized as follows. First we 
introduce the linear response theory for charge and heat currents
in unconventional superconductors.  Then we discuss in general the 
conditions necessary for observing the Wiedemann-Franz law deep
in the superconducting state.  Next, several order parameter
models are introduced and the low-temperature behavior of the 
corresponding electrical and thermal conductivities are discussed.
Finally, we compute the thermal conductivities
in the superconducting phase and compare our results
with recent experimental measurements in \UPt\ and 
$Zn$ doped \YBCO\ single crystals.

\section{Electrical and Thermal Conductivities}

We consider a superconductor with anisotropic singlet pairing or
unitary triplet pairing and discuss the electrical and thermal
conductivities in the long wavelength limit $q\!\to\! 0$, and at
temperatures $T\!\to\! 0$. For simplicity we assume isotropic impurity
scattering, which we treat self-consistently to leading order in $1/k_f\xi_0$.
In this case the first order corrections to the current
response functions of the impurity self-energy, $\delta\hat{\sigma}_{imp}$,
and the order parameter, $\delta\hat{\Delta}$, vanish for all pairing states
listed in Table~\ref{table:groups}.
Self-energy corrections corresponding to the excitation of collective 
modes of the order parameter, $\delta\hat\Delta$, also vanish in the 
limit $q\!\to\! 0$ (Hirschfeld \et\ 1989, Yip \et\ 1992).
The self-energy corrections are also called
`vertex corrections' in the Green function formulism of the Kubo 
response function (Rickayzen 1976).
If vertex corrections do not contribute, the response functions for 
the electrical and thermal conductivities depend only on the 
equilibrium propagators and self-energies, and the external 
perturbations.
A detailed derivation and justification of the linear response
in the quasiclassical formulation of a Fermi liquid has been 
given earlier by several authors and is repeated in brief
in the Appendix.
Hence we simply present the response functions and start 
with the discussion of our results for various order parameter models.  

  For spin singlet states the electrical conductivity is given by
\begin{eqnarray} \label{electrical}
&\displaystyle{
\mbox{Re\,}\sigma_{ij}(\omega,T)\:  = {e^2 N_{\!f}\hbar \over \pi
\omega}\int\!
  d\epsilon\int\! d\vec{p}_{\!f} \, v_{\!f, i}v_{\!f, j}\, 
\left[
  f\left({\epsilon_{-}}\right)- f\left({\epsilon_{+}}\right)
\right]}
\nonumber\\ &\times\, \displaystyle{
\mbox{Re\,}\Biggl\{ 
	M^R(\vec{p}_{\!f};\epsilon,\omega)
  \left(
	g_0^{R}(\vec{p}_{\!f};\epsilon_{-}) g_0^{R}(
	\vec{p}_{\!f};\epsilon_{+})
	+\underline{f}_0^{R}(\vec{p}_{\!f};\epsilon_{-})
	 f_0^{R}(\vec{p}_{\!f};\epsilon_{+}) +\pi^2
  \right)
}\nonumber\\
&\displaystyle{
\: - M^a(\vec{p}_{\!f};\epsilon,\omega)
  \left(
	g_0^{A}(\vec{p}_{\!f};\epsilon_{-}) g_0^{R}
	(\vec{p}_{\!f};\epsilon_{+})
	+\underline{f}_0^{A}(\vec{p}_{\!f};\epsilon_{-})
	 f_0^{R}(\vec{p}_{\!f};\epsilon_{+}) +\pi^2
  \right) \Biggr\}
}\,,
\end{eqnarray}
where $\epsilon_{\pm}=\epsilon\pm\hbar\omega/2$, and $f(\epsilon)$ is 
the Fermi-Dirac distribution function,
and $\int\! d\vec{p}_{\!f}\dots$ is a normalized Fermi surface integral.
For triplet pairing the spin scalar product is replaced by its
corresponding spin vector product,
$\underline{f_0} f_0 \to \underline{\vec f}_0 \cdot \vec f_0$.
The {\it retarded} ($R$) and {\it anomalous} ($a$) auxiliary functions
$M^X$ are defined as (for more details see the Appendix),
\begin{eqnarray}
M^X(\vec{p}_{\!f};\epsilon,\omega) &=& 
	{C^X_+(\vec{p}_{\!f};\epsilon,\omega) \over \pi^2
	  C^X_+(\vec{p}_{\!f};\epsilon,\omega)^2+
	  D^X_-(\epsilon,\omega)^2} \quad  \mbox{for}\ X\in\{R,a\} \,.
\end{eqnarray}  

  The result for $\sigma_{i j}(\omega,T)$ was obtained earlier
for electron-phonon and impurity scattering in conventional
superconductors in the strong coupling limit by Lee \et\ (1989)
and for the in-plane conductivity of layered weak-coupling superconductors 
by Graf \et\ (1995).  The conductivity formula reduces for a dirty 
$s$-wave superconductor to the well-known result of Mattis and Bardeen (1958).

In contrast to $\sigma_{ij}$, the thermal conductivity for a spin singlet state 
is determined solely by the anomalous part of the response function,
\begin{eqnarray}\label{thermal}
&&\displaystyle{
\kappa_{ij}(T)\: = {N_{\!f}\over \pi k_B T}\int\! d\epsilon
	\int\! d\vec{p}_{\!f} \, v_{\!f, i}v_{\!f, j}\, 
	\ \epsilon^2 \ {\frac{\partial f(\epsilon)}{\partial \epsilon}}
}\nonumber\\ &&
\hspace{12mm} \times\,
	M^a(\vec{p}_{\!f};\epsilon,0)
	\Big(
	g_0^{A}(\vec{p}_{\!f};\epsilon) g_0^{R}(\vec{p}_{\!f};\epsilon)
       -\underline{f}_0^{A}(\vec{p}_{\!f};\epsilon)
	f_0^{R}(\vec{p}_{\!f};\epsilon) +\pi^2 \Big) \,.
\end{eqnarray}
The retarded and advanced contributions dropped out 
after applying the normalization condition.
Physically, this means that the deviation of the
quasiparticle distribution function due to a thermal gradient
determines the heat current; changes in the quasiparticle
and Cooper pair spectrum do not. Equation (\ref{thermal}), combined
with the equilibrium propagators, impurity self-energy and order
parameter, is the basic result for the electronic contribution to the
thermal conductivity tensor.
It can be shown that $\kappa_{ij}$ in equation (\ref{thermal}) reduces 
to the same expression for the thermal conductivity as reported 
earlier by Schmitt-Rink {\it et al.} (1986), Hirschfeld {\it et
al.} (1986, 1988), and Fledderjohann \& Hirschfeld (1995), except that 
these authors do not include the $D^a_{-}$ term from the impurity self-energy.
The $D_{-}^X$ terms
are only significant at low temperatures,
where they can help to distinguish between magnetic and nonmagnetic
impurities,
and vanish in both Born and unitarity limits.

\subsection{The Wiedemann-Franz law}

In the limit $T\!\to\! 0$ and $\omega\!\to\! 0$ the occupation factors
$f(\epsilon{{\hspace*{-1pt}-\hspace*{-1pt}}} {\hbar\omega\over 2})-
f(\epsilon{{\hspace*{-1pt}+\hspace*{-1pt}}} {\hbar\omega\over 2})$ and
${\partial f}(\epsilon)/{\partial\epsilon}$ confine the 
$\epsilon$-integrals in (\ref{electrical}) and (\ref{thermal}) 
to a small $\epsilon$-region of order $k_B T$ or $\hbar\omega$.
  Assuming the existence of an energy scale $\epsilon^\ast \gg k_B T$,
on which the propagators and self-energies vary, we can set
$\epsilon=0$ in the slowly varying parts of the integrands.
  Using the normalization condition, $g_0^{R}(\vec{p}_{\!f})^2-
 \underline{f}_0^{R}(\vec{p}_{\!f})f_0^{R}(\vec{p}_{\!f}) =-\pi^2$,
in addition to the general symmetry relations for the Green functions,
we obtain for the conductivities
\begin{eqnarray} \label{electrical2}
&\displaystyle{
\mbox{Re\,}\sigma_{ij}(\omega\!\to\! 0, T\!\to\! 0)\:
= {2 e^2 N_{\!f} \hbar \over \pi }\int\!\! d\epsilon\!
\left( -\frac{\partial f(\epsilon)}{\partial \epsilon} \right)\!
\int\!\! d\vec{p}_{\!f}\, v_{\!f, i}v_{\!f, j}
\frac{ g_0^{R}(\vec{p}_{\!f})^2 }{ \pi^2 C^R_+(\vec{p}_{\!f}) }\,,
}
\end{eqnarray}
and
\begin{eqnarray}\label{thermal2}
&\displaystyle{
 \kappa_{ij}(T\!\to\! 0)\:
= {2 N_{\!f} \hbar \over \pi k_B T }\int\!\! d\epsilon
\,\epsilon^2 \!
\left( -\frac{\partial f(\epsilon)}{\partial \epsilon} \right)\!
\int\!\! d\vec{p}_{\!f}\, v_{\!f, i}v_{\!f, j}\:
 \frac{ g_0^{R}(\vec{p}_{\!f})^2 }{ \pi^2 C^R_+(\vec{p}_{\!f})}\,.
}
\end{eqnarray}
It is useful to  write our
final results in terms of a mean Fermi velocity and
an {\em effective} transport scattering time
$\tau_{ij}$, which incorporates all of the coherence effects of
superconductivity at $T\to 0$.

The energy integrals are standard, so the conductivities for a system
with $D$ dimensions reduce to
\noindent
\begin{eqnarray} \label{electrical3}
\mbox{Re\,}\sigma_{ij}(\omega\rightarrow 0, T\rightarrow 0)\: &=&
e^2 {2\over D} N_{\!f} v_{\!f,i} \tau_{ij} v_{\!f,j} \,,
\\
\label{thermal3}
\kappa_{ij}(T\rightarrow 0)\: &=&
{\pi^2 k_B^2\over 3}T \,{2\over D} N_{\!f} 
	v_{\!f,i} \tau_{ij} v_{\!f,j} \,,
\end{eqnarray}
where
$v_{\!f,i}^2 =  \int\! d\vec{p}_{\!f} \vec{v}_{\!f,i}(\vec{p}_{\!f})^2$,
and the effective transport time is defined by the tensor
\begin{eqnarray}\label{lifetime}
\tau_{ij} = -{D \hbar \over 2 v_{\!f,i}v_{\!f,j}}
\int\! d\vec{p}_{\!f}\,
{ v_{\!f, i}(\vec{p}_{\!f})v_{\!f, j}(\vec{p}_{\!f})\:
  \tilde\epsilon^R(0)^2
\over
  \Bigl(| \Delta(\vec{p}_{\!f}) |^2-
  \tilde\epsilon^R(0)^2
  \Bigr)^{3/2}
} \,.
\end{eqnarray}
Here $\tilde\epsilon^{R}(\epsilon)$ is the impurity renormalized 
quasiparticle excitation spectrum.  

  For an isotropic normal metal one has
$\tilde\epsilon^R(0)=i\hbar/2\tau$, where $\tau$ is the quasiparticle
lifetime due to impurity scattering in the normal state.  The transport
lifetime in the normal state reduces to $\tau$ for isotropic impurity
scattering, i.e.,\  $\tau_{ij}=\tau\delta_{ij}$.
  Note that expression (\ref{lifetime}) is applicable to the normal 
state because the key assumption in deriving 
(\ref{electrical3}) and (\ref{thermal3})
was that $T$ is small compared with $\epsilon^\ast$, where
$\epsilon^\ast$ is the energy on which the propagators and
self-energies vary.  Thus, for the normal state $\epsilon^\ast \sim
E_f$, while for the superconducting state $\epsilon^\ast \sim \gamma$,
where $\gamma$ is the impurity bandwidth of Andreev states,
defined by the transcendental equation
\begin{equation}
1 = \frac{n_i}{\pi N_{\!f}}
   \, 
   { \left< { 1 /\sqrt{ | \Delta(\vec{p}_{\!f}) |^2 + \gamma^2 }  } 
     \right>
    \over \cot^2 \delta_0 +  
    \left<
     { \gamma  /\sqrt{ | \Delta(\vec{p}_{\!f}) |^2 + \gamma^2 }  }
    \right>^2 } \,,
\label{gamma}
\end{equation}
with the Fermi surface average $\langle \dots \rangle$, the scattering
phase shift $\delta_0$, and the impurity concentration $n_i$.

  In some respect the band
of impurity states forms a new low-temperature metallic state deep in the
superconducting phase.  This analogy becomes more transparent 
when we calculate the temperature corrections to the transport
coefficients, using a Sommerfeld expansion.  However, the `metallic'
band of impurity states has other properties that differ significantly
from those of conventional metals.  The special features of the
impurity induced metallic band reflect the reduced dimensionality of
phase space for scattering and the energy dependence of the
particle-hole coherence factors.
Both features combined lead to (i) the universality of the transport 
coefficients at $T\to 0$ for excitation gaps with line nodes or quadratic
point nodes, and (ii) the temperature dependence of the Lorenz ratio
for elastic scattering at $T^\star < T < T_c$.

  We emphasize that (\ref{electrical3}) and (\ref{thermal3}) hold for
gapless superconductors in which the leading contribution to the
transport current is that from quasiparticle excitations with energies
$T \la T^\star \ll T_c$.  For superconductors with a gap at
the Fermi surface the number of quasiparticle excitations is of an
activated type at low temperatures, $\sim \exp{(-\Delta_0/T)}$, and
thus the transport coefficients cannot be described by equations
(\ref{electrical3}) and (\ref{thermal3}), with the consequence that
the Wiedemann-Franz law is strongly violated below $T_c$.

\subsection{Order parameter models}

  In our analysis we take a very simple approach.
Instead of examining the effects of a multi-sheet Fermi 
surface on the heat and charge current, we model the 
excitation spectrum by an 
excitation gap that opens at line and point nodes on the Fermi surface,
and by the Fermi surface properties in the vicinity of the nodes 
(\ie the Fermi velocities, $\vv_f$, and the density of states, $\Nf$,
near the nodes).  The low-temperature
behavior of the transport coefficients probes these `lower-dimensional'
regions of the Fermi surface, $\epsilon\ll\Delta_0\ll E_f$,
where the excitation gap vanishes, and is
less sensitive to the overall geometry of the Fermi surface.  It is
therefore admissible to parametrize the nodal regions of the gap with a
minimal set of nodal parameters.
The advantage of this approach is that we can quantitatively determine 
the phase space contributing to the low-temperature transport coefficients 
and then examine in more detail the
effects of impurity scattering and order parameter symmetry on the heat
current, without having to know the overall shape of the Fermi surface
or basis functions.

We consider models for the order parameter belonging to several
irreducible representations of the $D_{6h}$ and $D_{4h}$ symmetry groups,
which are chosen because they represent the primary candidates for the
pairing symmetry of the heavy fermion superconductor UPt$_3$
[for more details on the heat conduction at low temperatures
for the various order parameter models see Graf \et\ (1999)],
and the high T$_c$ superconductor \YBCO.\footnote{For
detailed discussions of the possible pairing models
in UPt$_3$ and high T$_c$ superconductors see the
reviews on \UPt\ by Sauls (1994a) and Heffner \& Norman (1996),
and on the cuprates by Pines (1994a) and Scalapino (1995).}
The singlet and triplet pairing states, we consider, have the form
\ber
({\rm singlet})\quad
\Delta(\vec{p}_{\!f}) &=& \Delta_0(T)\, 
	{\cal Y}_{\Gamma}(\vec{p}_{\!f}) \,,
\\
({\rm triplet})\quad
\vec\Delta(\vec{p}_{\!f}) &=& \Delta_0(T)\,
	{\cal Y}_{\Gamma}(\vec{p}_{\!f}) \,{\bf \hat d}\,,
\eer
where ${\bf \hat d}$ is the quantization axis along which the pairs have
zero spin projection.\footnote{
In this analysis spin has no effect on the transport coefficients 
beyond the connection between the direction of the spin quantization 
axis and the nodal structure of the odd-parity, triplet state basis 
functions.}

%
\begin{table}[h]
\caption{Symmetry groups, irreducible representations, and the standard
basis functions of the low temperature phase of several pairing models.}
\begin{flushleft}
\renewcommand{\arraystretch}{1.2}
\label{table:groups}
\begin{tabular}{ccccc}
\noalign{\smallskip}\hline\hline
    {Group}
  & {$\Gamma$}
  & {${\cal Y}_\Gamma$}
  & {point nodes}
  & {line nodes}
\\[1.0ex]
\noalign{\smallskip}\hline\noalign{\smallskip}
    $\rm D_{4h}$ & $\rm B_{1g}$ & $k^2_x - k^2_y$
  & --		& $\varphi_n=(2n+1)\, \frac{\pi}{4}, n=0,..,3$
\\[1.0ex]
    $\rm D_{6h}$ & $\rm E_{1g}$ & $k_z\,(k_x + i\, k_y)$
  & $\vartheta=0,\pi$	& $\vartheta=\frac{\pi}{2}$	
\\[1.0ex]
    $\rm D_{6h}$ & $\rm E_{2u}$ & ${\bf \hat c}\, k_z\,(k_x + i\, k_y)^2$
  & $\vartheta=0,\pi$	& $\vartheta=\frac{\pi}{2}$	
\\[1.0ex]
    $\rm D_{6h}$ & $\rm B_{1u}$ & ${\bf \hat c}\, \Im (k_x + i\, k_y)^3$
  & $\vartheta=0,\pi$	 & $\varphi_n=n\,\frac{\pi}{3},
    n=0,..,5$
\\[1.0ex]
    $\rm D_{6h}$ & $\rm A_{2u} \oplus i\, B_{1u}$
  & $\quad {\bf \hat c}\,[A\,k_z\,\Im (k_x + i\, k_y)^6$
  & $\vartheta=0,\pi$	 & $\varphi_n=n\,\frac{\pi}{3},
    n=0,..,5$
\\
	& & $\quad {+}{\it i}\,B\, \Im (k_x + i\, k_y)^3 ]$ & &
\\[1.0ex]
    $\rm D_{6h}$
  & $\rm A_{1g} \oplus i\, E_{1g}$
  & $\quad  A\,(2 k_z^2 - k_x^2 - k_y^2)$
  & cross-nodes:	 & $\vartheta = \cos^{-1}\frac{\pm 1}{\sqrt{3}}$
\\
	& & $\quad {+}{\it i}\,E\, k_y k_z ]\quad $ & & $\wedge\,
	\varphi = 0, \pi$
\\
\noalign{\smallskip}\hline\hline
\end{tabular}
\renewcommand{\arraystretch}{1.0}
\end{flushleft}
\end{table}

\begin{figure}
\begin{minipage}{0.49\textwidth}(a)
\centerline{ \epsfysize=55mm { \epsfbox{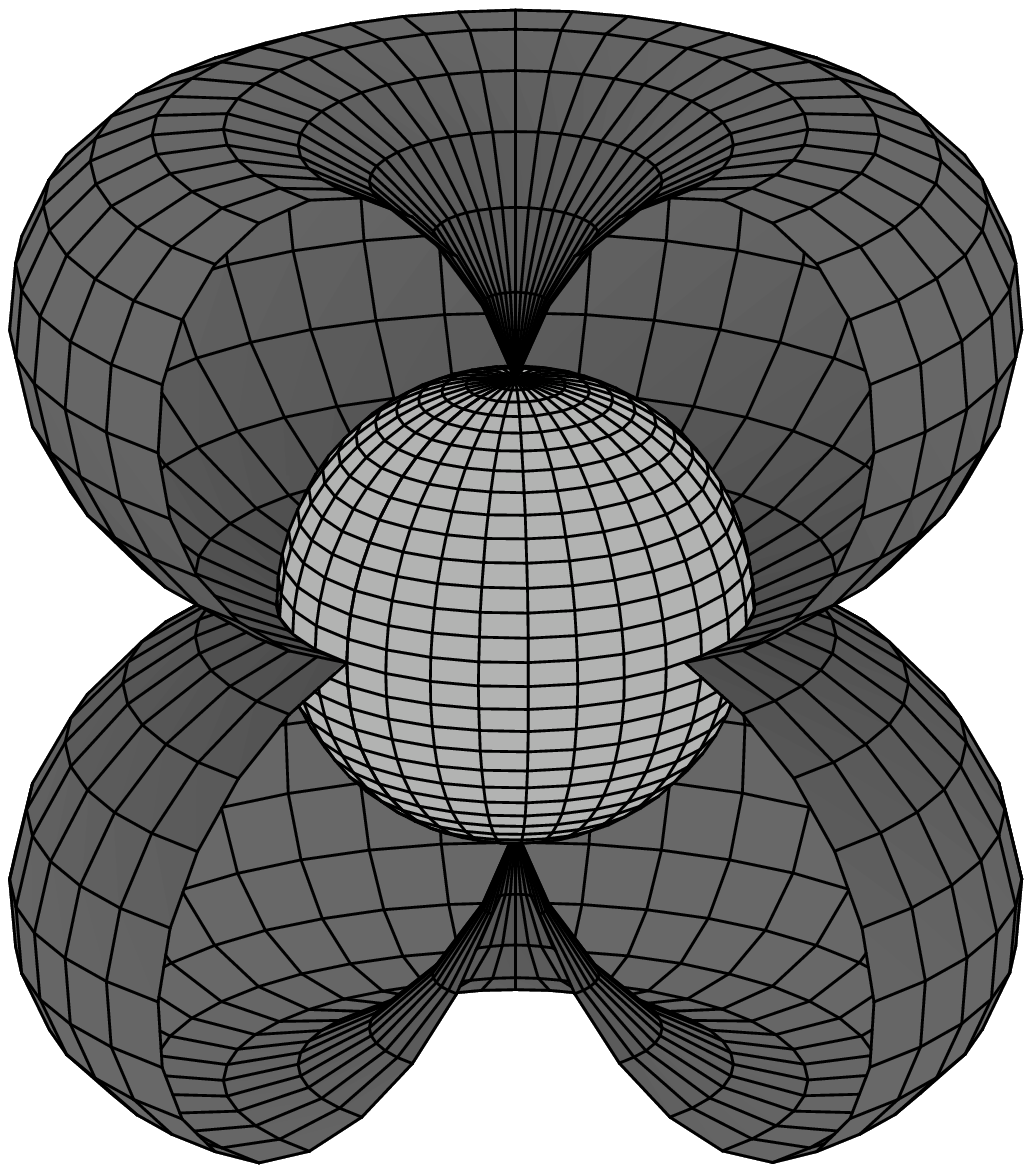} } }
\end{minipage}
\hfill
\begin{minipage}{0.49\textwidth}(b)
\centerline{ \epsfysize=55mm { \epsfbox{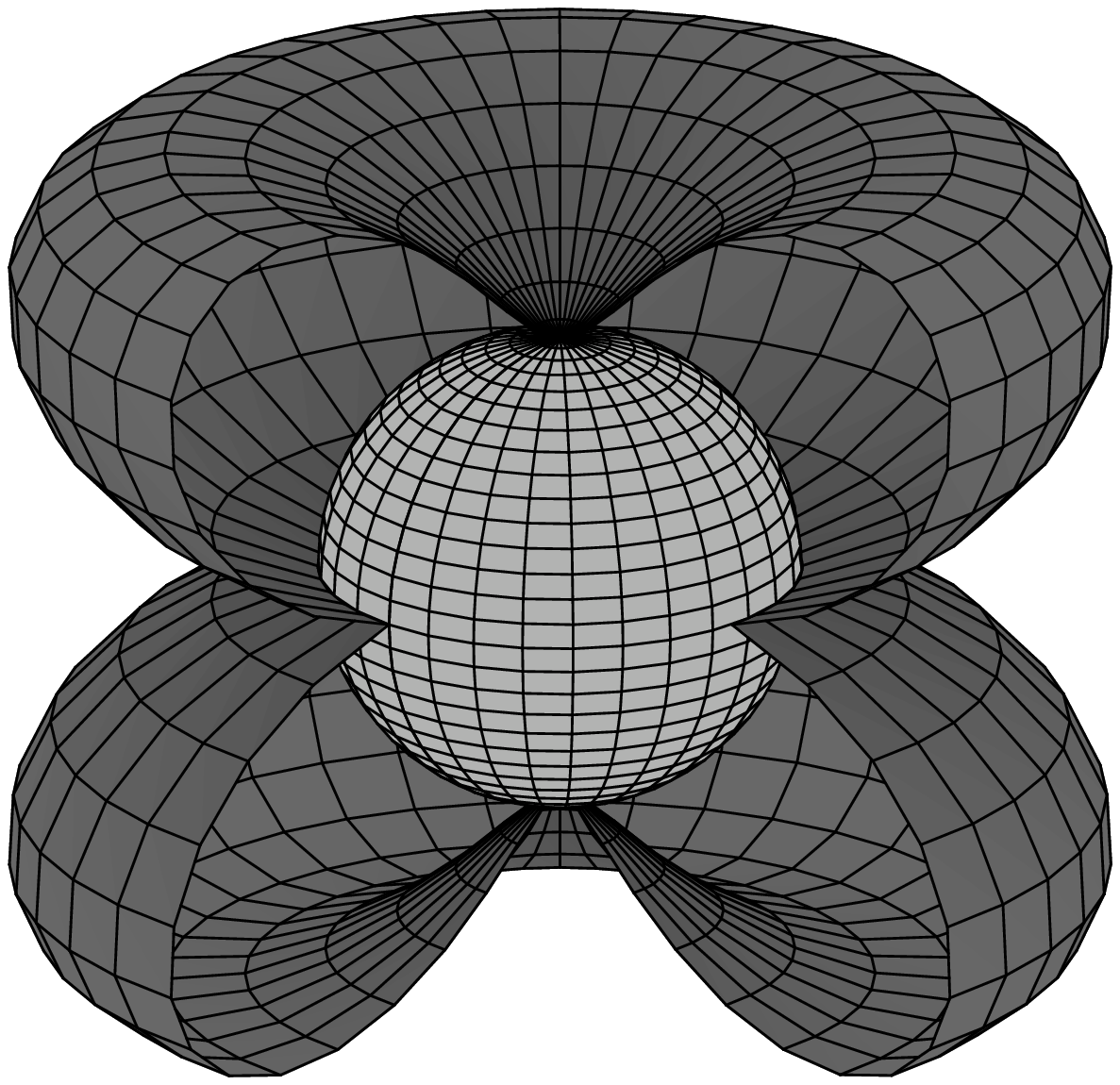} } }
\end{minipage}
\begin{minipage}{1.0\textwidth}(c)
\centerline{ \epsfxsize=45mm\rotate[r] { \epsfbox{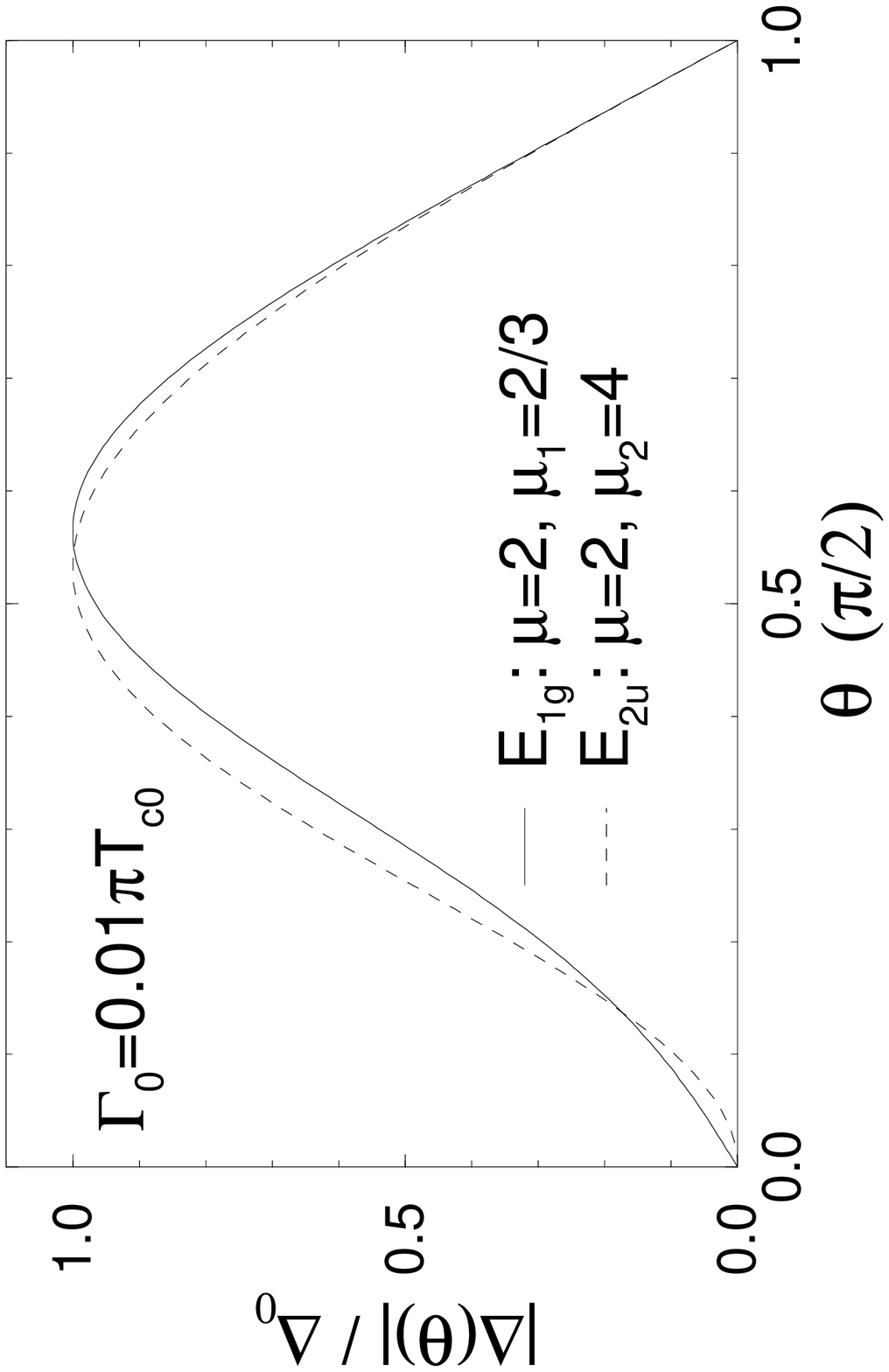} } }
\end{minipage}
\caption[]{
The excitation gap (exaggerated) at the Fermi surface (sphere) for
E$_{1g}$ (a) and  E$_{2u}$ (b) order parameters showing the line node
in the basal plane and the linear and quadratic point nodes at the
poles.  
Panel (c) shows the normalized excitation gap at $T=0$ for both states as
a function of the polar angle $\Theta$. The parameters correspond to
those that fit the low temperature thermal conductivity data shown in
Fig.~\ref{fig1}.
} \label{fig0}
\end{figure}

The basis functions ${\cal Y}_{\Gamma}$ listed in Table~\ref{table:groups}
have nodes in the order parameter and excitation 
gap that are dictated by symmetry. We model the different
nodal regions of the gap by a set of parameters, $\{\mu_i\}$
that determine the variation of the gap near the nodes (Graf \et\ 1996a).
Consider the E-rep models shown in Fig.~\ref{fig0}.
In the vicinity of the equatorial line node ($p_{\!fz}\!=\! 0$), 
$|\Delta(\theta)|\sim \mu_{\mbox{\scriptsize line}}\Delta_0|\Theta|$,
while near the poles ($p_{\!fx}\!=\! p_{\!fy}\!=\! 0$),
$|\Delta(\theta)|\sim \mu_{\mbox{\scriptsize
point}}\Delta_0|\Theta|^{n}$,
where the internal phase winding number is $n=1$ for E$_{1g}$ and 
$n=2$ for E$_{2u}$.\footnote{
  The parameters $\mu_{\mbox{\scriptsize line}}$ and
 $\mu_{\mbox{\scriptsize point}}$ define the slope or curvature of the
 gap near a nodal line or point in a spherical coordinate system that
 is obtained by mapping an ellipsoidal Fermi surface onto a sphere.}

%

\subsection{Universal limits and low temperature corrections}

  For simplicity we restrict our discussion of the universal limits and
low-$T$ corrections to the E-rep models proposed for \UPt, and list
the asymptotic values for the other pairing models in
Table~\ref{table:limits}.
  For a uniaxial, unconventional superconductor with an ellipsoidal 
Fermi surface a Sommerfeld expansion of the charge and heat 
conductivity tensors gives
\ber\label{Sommerfeld}
\left.
\begin{array}{lcl}
\sigma_i(T)   &\simeq& \alpha_{\sigma,i} + \beta_{\sigma,i}\; T^2\,,\\
\kappa_i(T)/T &\simeq& \alpha_{\kappa,i} + \beta_{\kappa,i}\; T^2\,,
\end{array}
\right.
\quad \mbox{for}\ i = a, b, c \,.
\eer
In the strong scattering limit the parameters $\alpha_b$ and $\beta_b$
are related to the microscopic model parameters by
\ber \label{universal_basal}
\alpha_{\kappa,b} \simeq \frac{v_{\!f,b}^2}{3}\,\gamma_S\, 
	\tau_{\scriptsize\Delta} \,, \qquad
\frac{\beta_{\kappa,b}}{\alpha_{\kappa,b}}  
\simeq \frac{7 \pi^2 k_B^2}{60\,\gamma^2} \,, \\
\alpha_{\sigma,b} \simeq \frac{3 e^2}{\pi^2 k_B^2}\,\alpha_{\kappa,b}
	\,, \qquad
\frac{\beta_{\sigma,b}}{\alpha_{\sigma,b}}  \simeq 
	\frac{5}{7}\frac{\beta_{\kappa,b}}{\alpha_{\kappa,b}}  \,.
\eer
Here $\gamma_S=\frac{2}{3}\pi^2 k^2_B N_{\!f}$
is the normal-state Sommerfeld coefficient, and the {\it effective}
transport scattering time is defined by
$\tau_{a}=\tau_{b}=\tau_{\scriptsize\Delta}$, where
\begin{equation}
\tau_{\scriptsize\Delta}=
3\hbar/(4\mu_{\mbox{\scriptsize line}}\Delta_0(0))
\,.
\end{equation}

  For heat flow along the $c$-axis the coefficients depend sensitively 
on the internal phase winding number, $n$,  of the order parameter 
around the point nodes,
\ber
\alpha_{\kappa,c} &\simeq& \frac{v_{\!f,c}^2}{3}\,\gamma_S
	\frac{3\hbar}{4\,\mu_{\mbox{\scriptsize point}} \Delta_0(0)} 
  \times
  \left\{
    \begin{array}{cc}
	\displaystyle
	{2\,\gamma}/{\mu_{\mbox{\scriptsize point}} \Delta_0(0)}\,, &
	({\rm E_{1g}})
	\\ \displaystyle
	{1} \,, & ({\rm E_{2u}})
   \end{array}
  \right.\, ,
\\
\alpha_{\sigma,c} &\simeq& \frac{3 e^2}{\pi^2 k_B^2}\alpha_{\kappa,c}\,,
\\
\frac{\beta_{\kappa,c}}{\alpha_{\kappa,c}}  &\simeq&
	\frac{7 \pi^2 k_B^2}{120\,\gamma^2}\times
  \left\{
    \begin{array}{cc}
	\displaystyle
	5\,, & ({\rm E_{1g}})
	\\ \displaystyle
	2\,, & ({\rm E_{2u}})
   \end{array}
  \right.\,, \qquad
\frac{\beta_{\sigma,c}}{\alpha_{\sigma,c}}  \simeq
	\frac{1}{7}\frac{\beta_{\kappa,b}}{\alpha_{\kappa,b}}\times
  \left\{
    \begin{array}{cc}
	\displaystyle
	3\,, & ({\rm E_{1g}})
	\\ \displaystyle
	5\,, & ({\rm E_{2u}})
   \end{array}
  \right.\,.
\label{caxis}
\eer
  One important result is that at low temperatures the
{\em basal plane} transport measurements cannot distinguish between the 
two pairing models, whereas $c$-axis transport coefficients are
sensitive to the symmetry of the pairing state. Small variations
in the concentration of defects may be used to probe the symmetry of the
pairing state.
  This is evident in the asymptotic slope of the thermal 
conductivity $\alpha_{\kappa,c}=\lim_{T\to 0}\kappa_c/T$, which
is universal for the E$_{2u}$ state, but depends on the impurity 
concentration ($n_i$) for the E$_{1g}$ state.  
  Furthermore, in the unitarity scattering limit
$\gamma^2\propto n_i$; thus, the coefficient of the $T^3$ term scales
with the impurity concentration as $\beta_c\sim 1/n_i$ for the
E$_{2u}$ state and $\beta_c\sim 1/\sqrt{n_i}$ for the E$_{1g}$ state.

The leading order finite temperature corrections
to the Wiedemann-Franz ratio become
\begin{equation} \label{WF}
L(T)  = { \kappa_b(T) \over T\,\mbox{Re\,}
\sigma_b(T,\omega\!\to\!0) }
\simeq
L_{\!S}\, \left( 1 + 
	\left[ \frac{\beta_\kappa}{\alpha_\kappa}-
	       \frac{\beta_\sigma}{\alpha_\sigma}
	\right]\,T^2
\right)
\simeq
L_{\!S}\, \left( 1 + {\pi^2 \over 30} {k_B^2 T^2 \over \gamma^2}
\right)
\,,
\end{equation}
which increases with temperature for $T\la T^*\sim\gamma/k_B$.  
This behavior arises from two sources:
(i) the residual density of states at $\epsilon=0$,
$N(0)\sim N_{\!f}\, (\gamma/\Delta_0)$,
and (ii) the difference in the coherence
factors between thermal and electrical conduction. Note
that for weak scattering, or very clean materials, the very
low-temperature regime $T<T^*$ may be difficult to achieve in practice.

The asymptotic limits for the order parameter models listed
in Table~1 are summarized in Table~2. Note the universal result
for the in-plane thermal conductivity of the $d_{x^2-y^2}$ model
for the high T$_c$ superconductors.
We omit the asymptotic value of the out-of-plane conductivity
in the cuprates because it strongly depends on the model for the interlayer 
transport (\ie coherent vs. incoherent propagation of quasiparticles).

%
\begin{table}[h]
\caption[]{Asymptotic values of the thermal conductivity tensor
$\tensor{\kappa}/T$ at $T\to 0$.}
\begin{center}
\renewcommand{\arraystretch}{1.2}
\label{table:limits}
\begin{tabular}{cccc}
\noalign{\smallskip}\hline\hline
    {Pairing State}
  & {Rep (Group)}
  & {$\displaystyle {\kappa_b(T)}
	\left(\frac{1}{2}\gamma_S {T}\,{v_{\!f,b}^2} \right)^{-1}$}
  & {$\displaystyle {\kappa_c(T)}
	\left(\frac{1}{2}\gamma_S {T}\,{v_{\!f,c}^2} \right)^{-1}$}
\\ \hline
    $d_{x^2-y^2}$
  & ${\rm B_{1g} \ (D_{4h})}$
  & $\displaystyle {2}/({\pi\mu\Delta_0})$
  & ---
\\[1.0ex]
    hybrid-I
  & ${\rm E_{1g} \ (D_{6h})}$
  & $\displaystyle {1}/({2\mu\Delta_0})$
  & $\displaystyle {\gamma}/({\mu_1^2\Delta_0^2})$
\\[1.0ex]
    hybrid-II
  & ${\rm E_{2u} \ (D_{6h})}$
  & $\displaystyle {1}/({2\mu\Delta_0})$
  & $\displaystyle {1}/({2\mu_2\Delta_0})$
\\[1.0ex]
    spin triplet
  & ${\rm B_{1u} \ (D_{6h})}$
  & $\displaystyle {3}/({2\mu\Delta_0)}$
  & $\displaystyle \sim\frac{1}{\mu_3\Delta_0}
	\left(	{\mu_3\Delta_0}/{\gamma} \right)^{1\over 3}$
\\[1.0ex]
    AB-model
  & ${\rm A_{2u}\! \oplus {\it i} B_{1u} \ (D_{6h})}$
  & $\displaystyle {3}/({2\mu\Delta_0})$
  & $\displaystyle \frac{ 
	\sim \left( \mu_3\Delta_0/\gamma \right)^{1\over 3}}{
	\mu_3\Delta_0} \ {\rm for}\ \mu_3 \sim 1$
\\[1.0ex]
    AE-model
  & ${\rm A_{1g}\! \oplus {\it i} E_{1g} \ (D_{6h})}$
  & $\displaystyle \frac{\sim \gamma^3}{
	(\mu_A\Delta_0^A)(\mu_E\Delta_0^E)^3}$
  & $\displaystyle \frac{\sim \gamma}{
	(\mu_A\Delta_0^A)(\mu_E\Delta_0^E)}$
\\
\noalign{\smallskip}\hline\hline
\end{tabular}
\renewcommand{\arraystretch}{1.0}
\end{center}
\end{table}
  
\begin{figure}
\centerline{ \epsfysize=80mm \rotate[r]{ \epsfbox{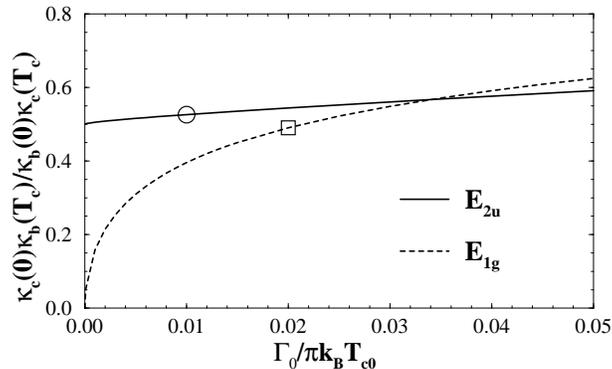} } }
\caption[]{
Impurity dependence of the thermal conductivity ratio for an
E$_{1g}$ and  E$_{2u}$ order parameter.  The symbols are the
low-temperature fits of the impurity concentration for the \UPt\
single crystal measured by Lussier \et\ (1996).
} \label{fig:impurity}
\end{figure}

  The low temperature anisotropy ratio $\kappa_c(T)/\kappa_b(T)$,
suggested as a symmetry probe by Fledderjohann and Hirschfeld (1995), 
distinguishes between the different E-rep pairing models and is very 
sensitive to the impurity concentration, see Table~\ref{table:limits}.  
  The experimentally observed anisotropy ratio in \UPt\ 
(Lussier \et\ 1996) approaches the value 
\beq
\lim_{T\to 0}\frac{\kappa_c(T)}{\kappa_b(T)} 
\approx 0.4\,\frac{\kappa_c(T_c)}{\kappa_b(T_c)} \,.
\eeq
  Assuming scattering in the unitarity limit and a reasonable 
value for the scattering rate 
$\Gamma(T_c) = \hbar/2\tau(T_c) \approx 0.06\dots 0.12\, k_B T_c$, 
we can account
for the data with
both pairing states \Eg\ and \Eu\ by adjusting 
the nodal parameters at the poles. 
We also can account for the data within the AE-model of 
Zhitomirsky \& Ueda (1996) 
by fitting the nodal parameters of the crossing line nodes
(cross-nodes).
  States with cubic point nodes, \eg the spin triplet model \Bu\ by
Machida and coworkers (1991, 1993) and  the accidental degeneracy AB-model
$\rm A_{2u}\oplus{\it i} B_{1u}$ by Chen \& Garg (1993, 1994),
have a much larger phase space contribution from the quasiparticles
scattering from impurities near the nodes.
  Further experiments on \UPt\ with controlled studies for various 
impurity concentrations should distinguish between 
the E-rep and AE models as shown for the E-rep models
in Fig.~\ref{fig:impurity}.

%

\section{Numerical results}

We explore the temperature dependence
of the thermal conductivity for unconventional superconductors.
Similar results were obtained earlier by several authors (see \eg\
Schmitt-Rink \et\ 1986, Hirschfeld \et\ 1986, 1988, 1994,
Monien \et\ 1987, 1988, Arfi \et\ 1988, 1989, Won \et\ 1993, and
Sun \et\ 1995).
What sets this work apart is the focus on the impurity induced band
of Andreev states, and their influence on the low-temperature
behavior of the heat current.  Here we extend our analytic 
analysis with numerical calculations of the low-$T$ behavior
beyond the universal region.
The numerical results are obtained by computing the equilibrium 
propagator, scattering self-energy, and order parameter self-consistently.
These results are then used to compute the transport coefficients.

\subsection{Thermal conductivity in \YBCO}

\begin{figure}
\centerline{
\epsfysize=90mm
\rotate[r] {\epsfbox{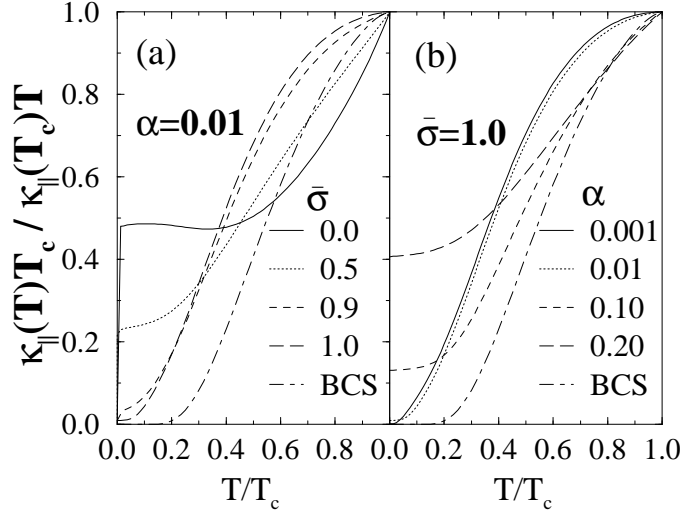}}}
\caption[]{
Thermal conductivity of a 2D $d_{x^2-y^2}$-wave superconductor.  
In panel (a): $\kappa/T$ for different scattering cross sections,
$\bar\sigma$, at a fixed scattering rate $\alpha=0.01$.
In panel (b): $\kappa/T$ for different elastic scattering rates 
$\alpha$ in the unitarity limit ($\bar\sigma=1.0$).  
The isotropic $s$-wave BCS result is shown for comparison.
}
\label{fig:kcond}
\end{figure}

In Fig.~\ref{fig:kcond} we plot for a $d_{x^2-y^2}$-wave pairing state,
$\Delta(\vec{p}_f) = \Delta_0\, \cos\, 2\phi$,
the thermal conductivity for several 
normalized scattering cross sections $\bar\sigma=\sin^2\delta_0$,
and normalized scattering rates $\alpha=\hbar/(2\pi k_B T_{c0}\tau)$.
\footnote{
Here we use simultaneously the notation of the dimensionless
parameter $\alpha$ and $\Gamma = \hbar/2\tau$ for the scattering
rate.}
A consequence of the universal limit is that the ratio
\begin{eqnarray}
\lim_{T\to 0} \frac{\kappa(T)\, T_c}{\kappa(T_c)\, T} \simeq
\frac{2\hbar}{\pi \tau \mu_{\mbox{\rm\scriptsize line}}\Delta_0(0)} 
=\frac{4\,\Gamma}{\pi\mu_{\mbox{\rm\scriptsize line}} \Delta_0(0)}
\end{eqnarray}
scales linearly with the scattering rate, $\Gamma$,
and is independent of the scattering strength $\bar\sigma$.
For $T \agt T^\star$ Arfi {\it et al.}\ (1989) have shown that
$\kappa(T)/T\propto (1-\bar\sigma)\,\kappa(T_c)/ T_c$
strongly depends on the scattering phase shift.
This explains (i) the sudden drop of
$[\kappa(T)\, T_c]/[\kappa(T_c)\, T]$ in Fig.~\ref{fig:kcond}(a)
at ultra-low temperatures for weak scattering, where the universal
limit is achieved only for 
$T\alt T^\star\sim (2\mu_{\mbox{\scriptsize line}}\Delta_0/k_B) 
\exp{(-\pi\mu_{\mbox{\scriptsize line}}\Delta_0/4\Gamma)}$,
and (ii) the scaling of the zero temperature intercept with $\alpha$ in
Fig.~\ref{fig:kcond}(b).
Below $T^\star$ the ratio $\kappa/T$ approaches the universal limit.
Weak scattering leads to an approximately linear temperature
dependence over a large portion of the temperature range
(Pethick \et\ 1986, Barash \et\ 1996).  However, 
$\kappa/T$ changes drastically in rather clean superconductors 
below the exponentially small crossover temperature, where it 
approaches its linear low-temperature asymptote.

Our calculations of the effects of the impurities on
the low-$T$ thermal conductivity assume a cylindrical Fermi
surface and purely elastic scattering processes. These
simplifications don't affect the low-temperature thermal conductivity
since in this limit it is governed by the nodal points.
However, we do not expect an accurate
description of the experimental data at higher temperatures,
$T \sim T_c$, where inelastic scattering is significant and excitations
over the whole Fermi surface contribute to the transport properties.

%

Recently Taillefer \et\ (1997) reported measurements of the
heat conductivity in $Zn$ doped 
YBa$_2$(Zn$_x$Cu$_{1-x}$)$_3$O$_{6.9}$.
The in-plane thermal conductivity was measured at temperatures as low as 
$T_c/1000$, and the predicted universal behavior of $\kappa/T$
was observed for the first time. They found that the asymptotic value of 
$\kappa/T$ is almost independent of the $Zn$ concentration ($0\%-3\%$),
with $\lim_{T \to 0}\, \kappa/T \approx 18\, {\rm mW/K^2 m}$.
They also found that phonons scattering off the crystal surfaces
cannot be neglected in \YBCO\ even at $T \sim T_c/1000$.
The reported asymptotic value is somewhat larger than our previous 
estimate of $\kappa/T \approx 14 \, {\rm mW/K^2 m}$ (Graf \et\ 1996b)
which is based on the expression
\beq
\lim_{T \to 0} \kappa/T
	= \frac{\pi^2}{3}\frac{k_B^2}{e^2} \frac{\omega_p^2
	\tau_{\scriptsize \Delta}}{4\pi}
	= \frac{1}{2} \gamma_S v_f^2 \tau_{\scriptsize \Delta} \,,
\eeq
with a Drude plasma frequency $\omega_p \approx 1.5\, \mbox{eV}$,
and $\tau_{\scriptsize \Delta} = {\hbar}/{\pi\Delta_0(0)}$,
but is well within the uncertainties of the known material parameters.
For example, if one uses specific heat measurements instead 
of the Drude plasma frequency, with typical values for the 
normal-state specific heat
$C_{\rm el}/T = \gamma_S \approx 270\pm 70\, {\rm J\,K^{-2} m^{-3}}$
(Junod \et\ 1990), and the effective Fermi velocity near 
the nodal regions, $v_f \approx 110\pm 20\,\mbox{km/s}$,
then we obtain $\kappa/T \approx 20 \, {\rm mW/K^2 m}$.  

\subsection{Thermal conductivity in \UPt}

The inelastic scattering rate in \UPt\ is comparable to 
elastic scattering at $T_c$ and dominated by electron-electron
scattering in the normal state for $0.5\,{\rm K} < T < 2\,{\rm K}$
(Lussier \et\ 1995).  Elastic scattering is roughly isotropic 
($s$-wave scattering).  However, recent resistivity measurements by 
Kycia \et\ (1998) show deviations of order $20\%$ from isotropic scattering.
  In order to describe the temperature dependence of the thermal 
conductivity we model elastic scattering, $\Gamma_0$, and inelastic 
scattering, $\Gamma_{\rm in}$,  by a phenomenological
scattering rate $\Gamma(T) \equiv \hbar/2\tau(T) = \Gamma_0 (1 + T^2/T_c^2)$,
where we neglect the effects of anisotropic scattering.
  The inelastic scattering rate decreases below $T_c$ even faster 
than $T^2$ because of the onset of superconductivity and the opening of 
a gap over most of the Fermi surface.  We will neglect this reduction 
in $\Gamma(T)$, since we are mostly interested in the low-temperature
regime, \eg  for temperatures $T \alt T_c/10$ the inelastic scattering
rate is less than $1\%$ and can safely be neglected.
However, we do not expect an accurate description of the experimental 
data in the temperature regime, $T_c/2\alt T \le T_c$, where inelastic 
scattering is not negligible.

  Our analysis of the universal regime implies rather stringent 
restrictions on an equatorial line node
\begin{eqnarray} \label{ratio_basal}
\lim_{T\to 0} \frac{\kappa_b(T)T_c}{\kappa_b(T_c)T} \simeq
        \frac{3}{2}\frac{\Gamma(T_c)}
        {\mu_{\mbox{\scriptsize line}}\Delta_0(0)}\,,
\qquad
k_B T^\star \simeq  0.2\, \sqrt{\mu_{\mbox{\scriptsize line}}
        \Delta_0(0)\, \Gamma_0} \,.
\end{eqnarray}
  The experimentally imposed constraints are consistent with a
nodal parameter $\mu_{\mbox{\scriptsize line}}=2$, an
intercept of $[\kappa_b(T)\,T_c] / [\kappa_b(T_c)\,T] \alt 0.02$
at $T \to 0$, and a crossover temperature $T^\star \alt T_c/12$
(Lussier \et\ 1996, Huxley \et\ 1996, Suderow \et\ 1997).
  The elastic scattering rate $\Gamma_0$ is adjusted for a best fit 
to the low-$T$ region for the basal plane thermal conductivity.
For an E$_{2u}$ pairing state a good fit is obtained for 
$\Gamma_0 = 0.01\pi\,T_{c0}\approx 0.03\,T_c$,
for which the theoretically computed intercept is 
\beq
\lim_{T\to 0} = \frac{\kappa_b(T)/T}{\kappa_b(T_c)/T_c}\approx 0.02\,.
\eeq
This scattering rate fits the low-temperature part of the 
thermal conductivity in the superconducting state, but is smaller than 
that estimated from the normal-state transport data and de Haas-van Alphen 
data, $\Gamma_0\approx 0.1\,T_c - 0.2\,T_c$, (Taillefer \et\ 1987).
Larger discrepancies were reported earlier
by Lussier \et\ (1995, 1996), and Fledderjohann \et\ (1995).
  Finally, the nodal parameters defining the excitation gap near the 
point nodes are adjusted to fit the heat current along the $c$-axis,
$\kappa_c(T)/T$ at $T\ll T_c$. 
These values are $\mu_{1} = 2/3$ for the E$_{1g}$ model and 
$\mu_{2} = 4$ for the E$_{2u}$ model.

\begin{figure}
\begin{minipage}{0.49\textwidth}
%
%
(a)\vspace*{-5mm}
\centerline{ 
\epsfysize=0.95\hsize\rotate[r] { \epsfbox{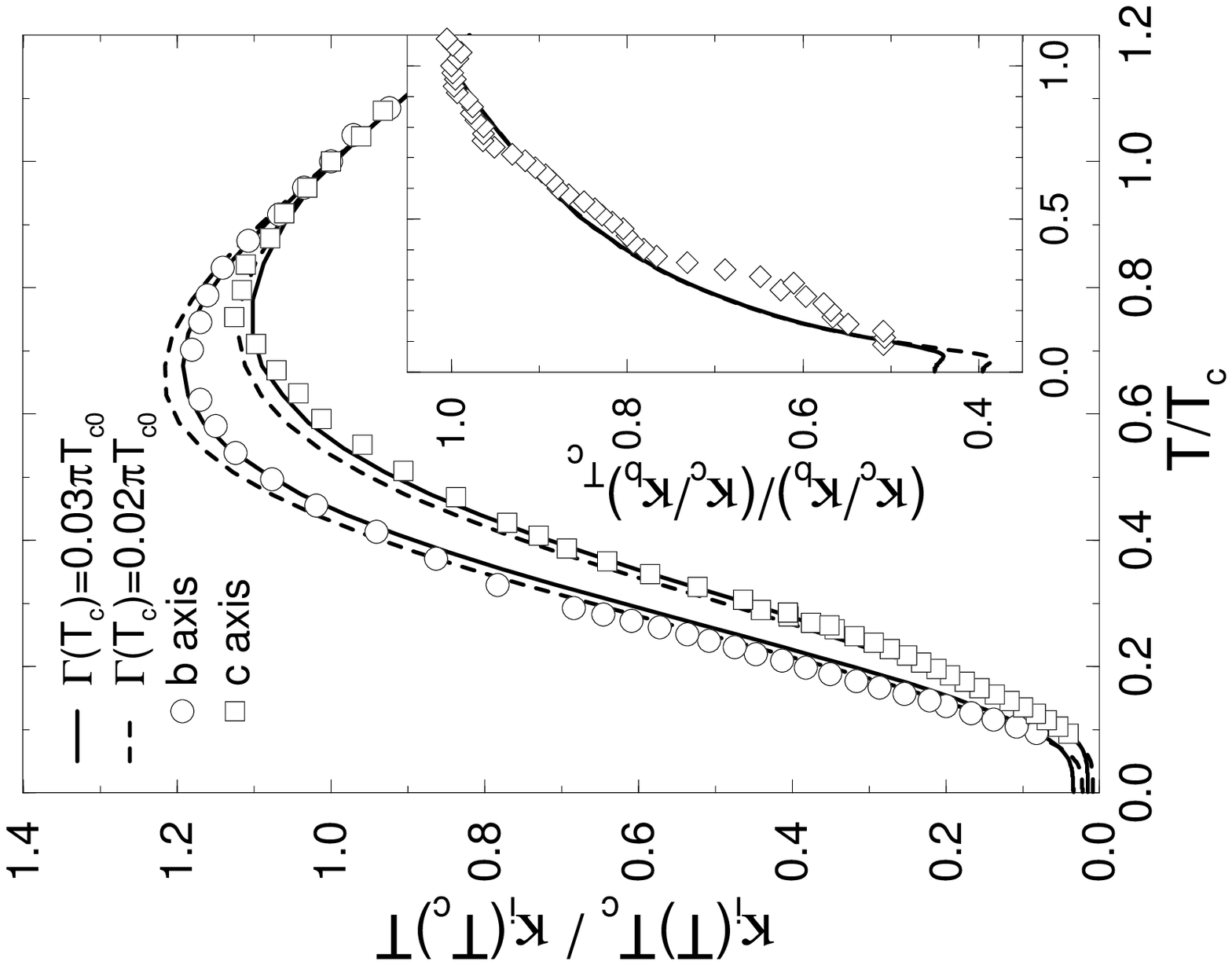} } 
}
\end{minipage}
\hfill
\begin{minipage}{0.49\textwidth}
%
%
(b)\vspace*{-5mm}
\centerline{ 
\epsfysize=0.95\hsize\rotate[r]{ \epsfbox{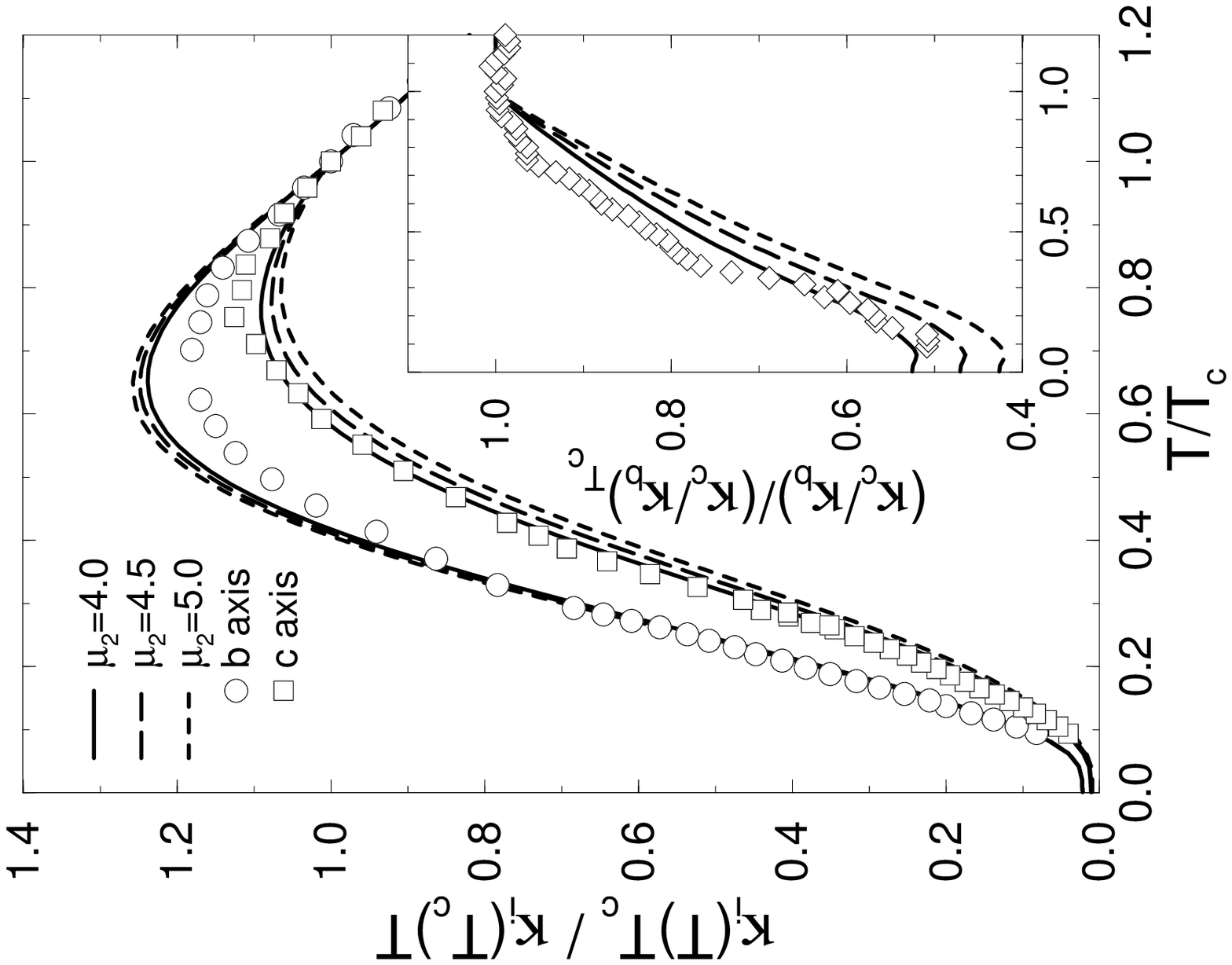} } 
}
\end{minipage}
\caption[]{
  Heat flow along the $b$ and $c$ axes in the unitarity limit.
The experimental data (symbols) were taken from Lussier \et\ (1996)
and normalized to the values at $T_c\simeq 0.5\,{\rm K}$.
The insets show the normalized anisotropy ratios.
Panel (a): E$_{1g}$-state for two scattering rates $\Gamma(T_c)$.
 The slope parameters at the linear point nodes and at the line node 
were fixed to $\mu_{\scriptsize 1}=2/3$ and 
$\mu_{\mbox{\scriptsize l}}=2.0$.
Panel (b): E$_{2u}$-state with a scattering rate 
$\Gamma_0=0.01\pi\,k_B T_{c0}$
and different nodal parameters at the point nodes,
$\mu_{\scriptsize 2}=4.0, 4.5, 5.0$. 
The slope of the line node was fixed to
$\mu_{\mbox{\scriptsize l}}=2.0$.
}\label{fig1} 
\end{figure}

\begin{figure}
\begin{minipage}{0.49\textwidth}
\centerline{ 
\epsfxsize=0.98\hsize
{ \epsfbox{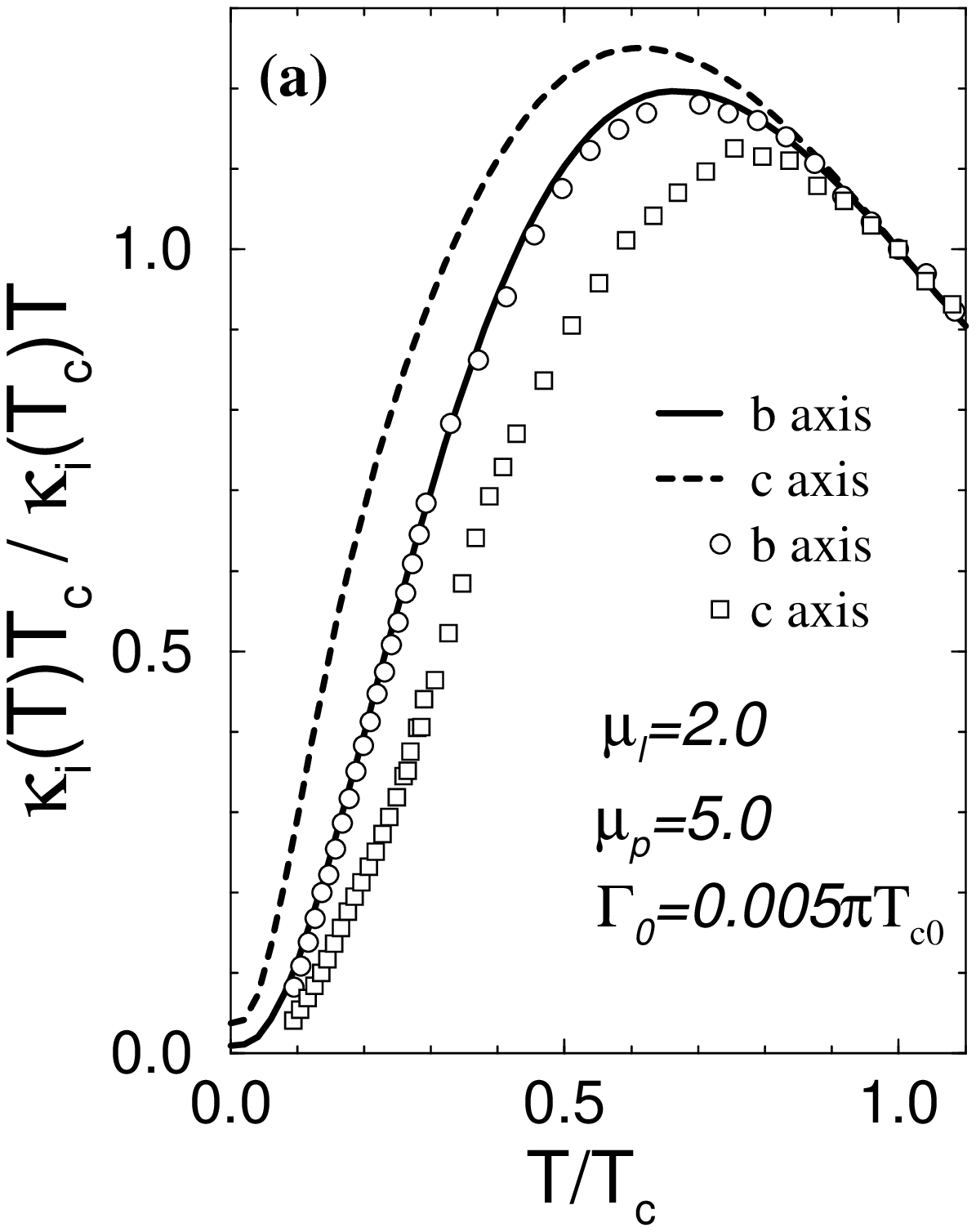} }
}
\end{minipage}
\hfill
\begin{minipage}{0.49\textwidth}
%
%
\centerline{ 
\epsfxsize=0.98\hsize
{ \epsfbox{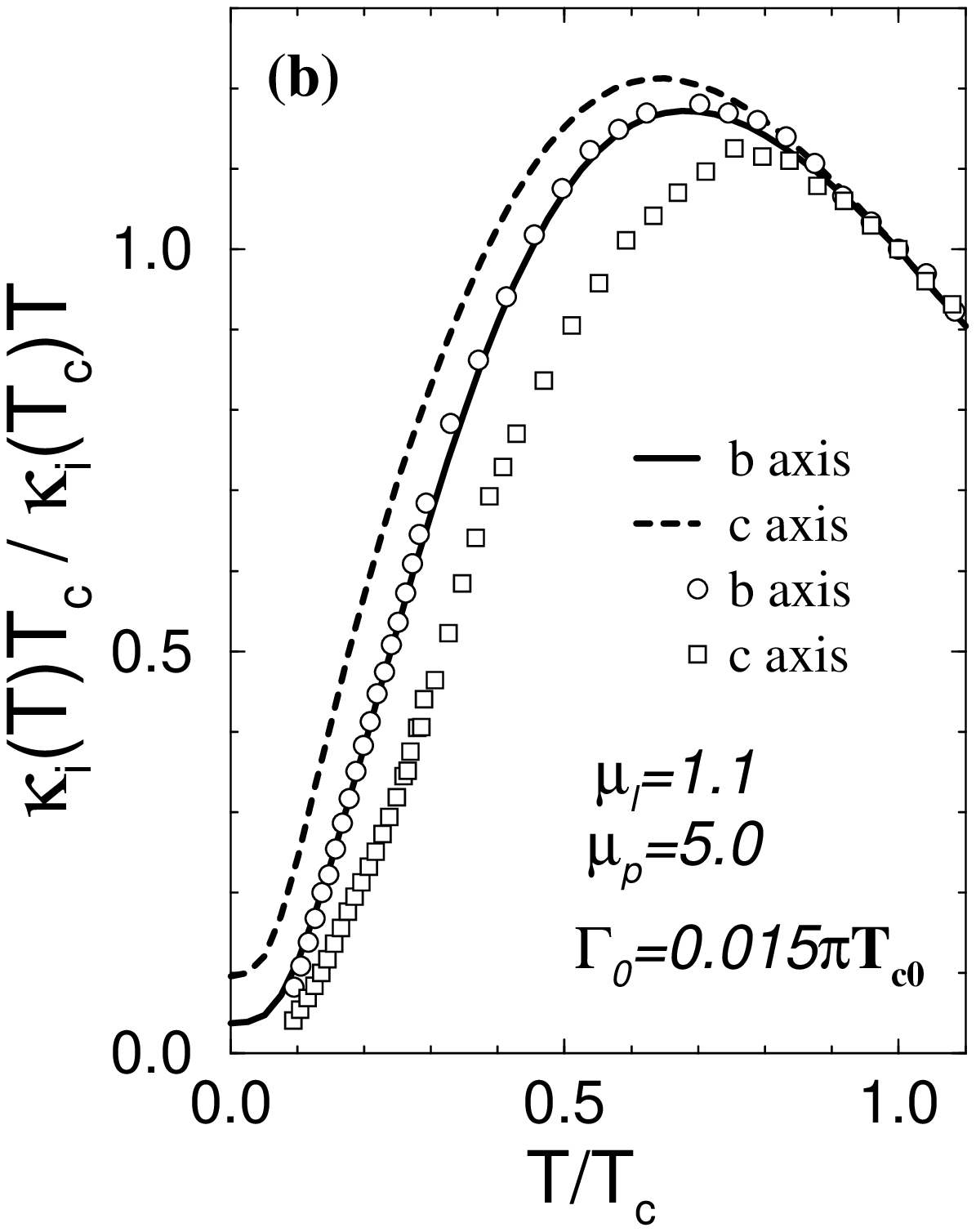} }
} 
\end{minipage}
\caption[]{
The thermal conductivity for heat flow along the $b$ and $c$ axes
in the resonant impurity scattering limit. The material parameters are
the scattering rate, $\Gamma_0$, and the nodal parameters, $\mu_i$,
at the point ($p$) and line ($l$) nodes.
The theoritcal results are compared with the experimental data 
(symbols) for \UPt.
Panel (a): fit for an odd-parity B$_{1u}$-state.
Panel (b): fit for an accidentally degenerate 
$A_{2u} \oplus i B_{1u}$-state.}
\label{fig:B1u}
\end{figure}

A similar analysis for two of the 
other proposed pairing models 
for \UPt\ is shown in Fig.~\ref{fig:B1u}.
It is in principal possible to fit the basal plane
thermal conductivity, however, a fit for the heat current along the 
$c$-axis generally fails because of the large phase
space contribution from the 
cubic point nodes and cross-nodes (crossing line nodes at the poles)
for the $B_{1u}$ and $A_{2u}\oplus iB_{1u}$ order parameters.
This is the case for all order parameter models other than the
E-rep models and the AE-model. 
Thus, the only models that are capable of explaining the
existing data for the anisotropy and temperature dependence of the 
heat conductivity in UPt$_3$ are the 2D E-rep models and the AE-model.
  
%

\begin{figure}
\centerline{ 
\epsfysize=90mm \rotate[r]{ \epsfbox{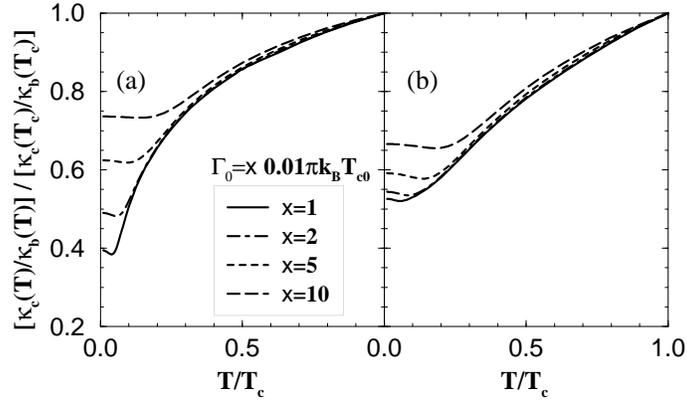} }
}
\caption[]{
  Thermal conductivity ratio for a set of
impurity scattering rates $\Gamma_0$ and a fixed 
$\Gamma_{\rm in} = 0.01\,\pi k_B T_{c0}\,(T/T_c)^2$
for an E$_{1g}$ (a) and E$_{2u}$ (b) pairing state. 
} \label{fig:kappa-ratio}
\end{figure}

The existing low-temperature data for the thermal conductivities of 
UPt$_3$ (Lussier \et\ 1996) are equally well described by either an 
E$_{1g}$ or E$_{2u}$ 
or AE model. 
However one possibility to distinguish between 
these models is by studying the anisotropy ratio $\kappa_c/\kappa_b$ 
in the low-temperature region for different scattering rates.
The corresponding zero temperature limit for the E$_{2u}$ (hybrid-II) 
state is universal, \ie\ independent of impurity scattering for 
$\gamma \ll \Delta_0$. It depends only on the nodal (material) 
parameters through the ratio
$\mu_{\mbox{\scriptsize line}}/\mu_{\mbox{\scriptsize point}}$.
In contrast the anisotropy ratio for the E$_{1g}$ state 
and AE pairing state depends
sensitively on the scattering rate and scattering 
strength, see Fig.~\ref{fig:kappa-ratio}.

%

  We summarize the quality of our fits of the thermal conductivity
and the observable differences between the various models for the
pairing state in \UPt\ by plotting the anisotropy ratio 
${\cal R}(T) = [\kappa_c(T)/\kappa_b(T)] / [\kappa_c(T_c)/\kappa_b(T_c)]$.
Figure~7 shows that the ratio ${\cal R}(T)$ for the data of Lussier 
\et\ (1996)
is in agreement with the pairing states for the \Eg, \Eu, and AE models.
The spin triplet model \Bu\ and the AB-model cannot account for the anisotropy
at low temperatures, ${\cal R} < 1$.  Independent measurements of the heat
currents by Suderow \et\ (1997) differ to some extend from those by
Lussier \et\ (1996), particularly in the ratio ${\cal R}$.  This discrepancy is
mainly due to differences in the heat current along the $c$-axis.
This may be caused by crystal imperfections along the crystal $c$-axis, and
needs further investigation.

%
\begin{figure}
\centerline{ 
\epsfysize=0.78\textwidth \rotate[r]{ \epsfbox{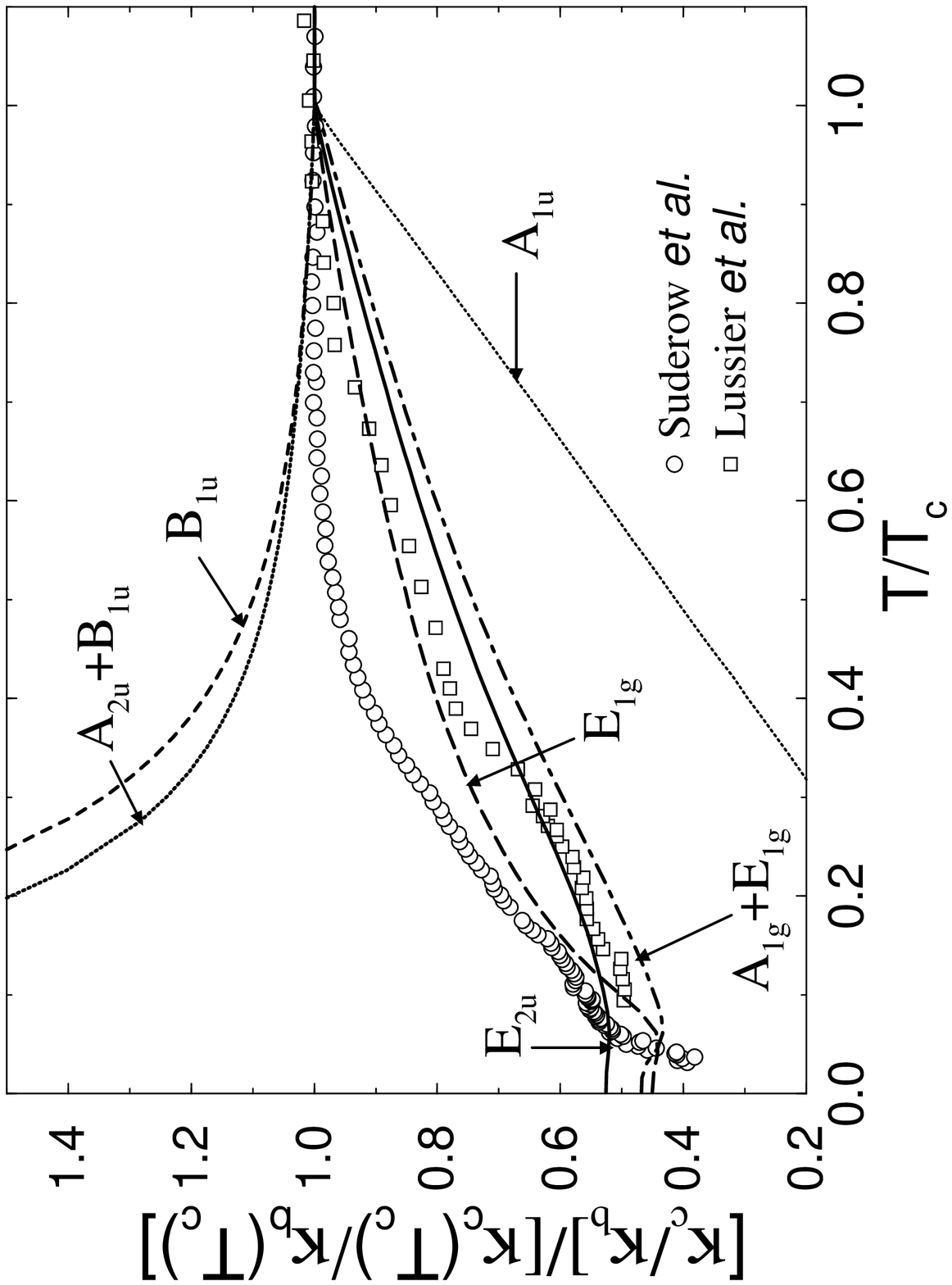} } 
}
\caption[]{
 The anisotropy ratio of the thermal conductivities for
various `best fit' pairing states. The line for the polar state
A$_{1u}$ is shown for comparison.
The experimental data are renormalized relative to the upper transition
($T_c^+$).
}\label{fig:R-ratio}
\end{figure}

\section{Conclusion}

  At very low temperatures heat transport is dominated by elastic 
scattering of quasiparticles in lower dimensional regions of the Fermi 
surface in the vicinity of the nodes of the order parameter. As a result
the low energy spectrum can be determined accurately without complete 
knowledge of the Fermi surface.
The residual density of states at zero energy and zero temperature
leads to a Wiedemann-Franz law for electronic transport
deep in the superconducting phase. The results for the low-temperature
transport coefficients can shed new light on the strength of the scatterering
centers because of the sensitivity of the coherence factors
to the scattering phase shift.
Furthermore, our theoretical results suggest that 
measurements of the heat transport along different crystal axes can
differentiate between the various pairing models that have been
proposed for heavy fermion and other unconventional superconductors.

  Excellent agreement with experiments on \UPt\ is obtained for pairing 
states with E$_{1g}$ or E$_{2u}$ symmetry at low temperatures,
as well as for the AE-model,
whereas no agreement (along the $c$-axis) could be obtained for the
paring states with \Bu\ and $\rm A_{2u}\oplus {\it i} B_{1u}$ symmetries.
The differences in the
excitation spectrum for \Eg\ and \Eu\ can be observed at
very low temperatures, $T\la T^\star \sim T_c/12$, when the
impurity-induced Andreev band determines the heat conductivity.
Measurements of the thermal conductivity ratio, $\kappa_c/\kappa_b$, at
very low temperatures ($T\ll T_c$) for various 
impurity concentrations should further
differentiate between these models; the E$_{2u}$ model is predicted
to have a universal ratio, while the ratio for the E$_{1g}$ state
depends strongly on the concentration and scattering strength of impurities.
At high temperatures, $T\sim T_c$, our calculations show deviations
from the experimental data, which reflect our neglect of
(i) the effect of the onset of superconductivity on the inelastic
scattering rate, and (ii) the splitting of the superconducting
transition temperature. Finally, we note that our prediction of
a universal $\kappa/T$ at low temperatures is in good agreement with
recent experiments on $Zn$ doped \YBCO\ (Taillefer \et 1997).

\section*{Acknowledgments}

This work was started at Northwestern University and supported 
by the National Science Foundation (DMR-9705473), the Science 
and Technology Center for Superconductivity (DMR 91-20000),
the Deutsche Forschungsgemeinschaft, and was completed at
Los Alamos National Laboratory under the auspices of the U.S.\ 
Department of Energy.

\appendix

\section{Linear Response in the Quasiclassical Formulism}

\subsection{Quasiclassical transport coefficients}

The electrical current density is determined by
the Keldysh propagator, the Fermi velocity, $\vec v_{\!f}$, and
the density of states per spin, $N_{\!f}$, at the Fermi level,
\begin{equation}\label{current}
\vec j_{e}(\vec R,t)=
	2N_{\!f}\int d\vec p_{\!f}\int{d\epsilon\over4\pi i}
	\big( e\,\vec v_{\!f}(\vec p_{\!f}) \big)\,
	\frac{1}{4} {\rm Tr}\,
	\hat\tau_3 \hat g^K(\vec p_{\!f},\vec R;\epsilon,t)\,,
\end{equation}
and the heat current density has the form
\begin{equation}\label{currenth} 
\vec j_{\varepsilon}(\vec R,t) =
	2N_{\!f}\int d\vec p_{\!f}\int{d\epsilon\over4\pi i} \,
	\big(
	  \epsilon\, \vec v_{\!f}(\vec p_{\!f}) 
	\big)	\,
	\frac{1}{4} {\rm Tr}\,
	\hat g^K(\vec p_{\!f},\vec R;\epsilon,t) \,.
\end{equation}

For weak disturbances from equilibrium the current response is linear
in the applied perturbation.  Here, we are interested in the
electrical and heat current response functions, which define the
conductivity tensors by
\begin{equation}
\delta\vec{j}_{e}=\vec{\sigma}\cdot\vec{E}_{\omega} 
\,, 
\quad\mbox{and}\qquad
\delta\vec{j}_{\varepsilon}=-\vec{\kappa}\cdot\vec{\nabla}T
\,.
\end{equation}
We develop the quasiclassical linear response equations
for the transport coefficients $\vec{\kappa}$ and $\vec{\sigma}$,
following the discussion by Rainer and Sauls (1995), 
and Graf \et\ (1995).

The advanced, retarded, and Keldysh propagators are
calculated from the quasiclassical transport equations
\begin{equation}\label{transp1}
\left[	\epsilon\hat\tau_3-\hat{\sigma}_{ext}-
	\hat\sigma^{R,A}\,,\, \hat g^{R,A}
\right]_{\circ} = 
\vec v_{\!f}\cdot\frac{\hbar}{i}\vec\nabla \hat g^{R,A}\,,
\end{equation}
and
\begin{eqnarray}\label{transp2}
\left(
  \epsilon\hat\tau_3-\hat{\sigma}_{ext}- \hat\sigma^{R}
\right)
\circ \hat g^{K}
-\hat g^{K}\circ
\left(
  \epsilon\hat\tau_3-\hat{\sigma}_{ext}- \hat\sigma^{A}
\right) \nonumber\\
-\hat\sigma^{K}\circ \hat g^{A} + \hat g^{R}\circ \hat\sigma^{K}
= \vec v_{\!f}\cdot\frac{\hbar}{i}\vec\nabla \hat g^{K}\,,
\end{eqnarray}
which are supplemented by the
normalization conditions (Eilenberger 1968, Larkin \& Ovchinnikov 1968),
\begin{eqnarray}\label{norm}
\hat g^{R,A} \circ \hat g^{R,A} =-\pi^2\hat 1 \enspace, 
\quad\mbox{and}\qquad
\hat g^R \circ \hat g^K + \hat g^K \circ \hat g^A =0 \enspace.
\end{eqnarray}

  To linear order in the perturbation the quasiparticles (with charge $e$) 
respond to an electric field or temperature gradient through the self-energy
terms,
\begin{equation}
\hat{\sigma}_{ext}=-\frac{e}{c}\,\vec{v}_{\!f}\cdot\vec{A}\,\hat{\tau}_3
\,,
\quad\mbox{or}\qquad
\vec v_{\!f}\cdot\frac{\hbar}{i}\vec\nabla \hat g^{X}_0
\,,
\quad
\mbox{for}\ X \in \{ R, A, K \}
\,,
\end{equation}
where $\vec{A}(\vec{q},\omega)$ is the vector potential describing
the transverse field $\vec{E}_\omega = ({i\omega}/{c})\vec{A}$,
and $\hat g^X_0$ are the equilibrium propagators. In order to
calculate the conductivity we solve the transport equations to
linear order in the perturbing fields. For the heat conductivity
the perturbation is the temperature gradient $\vec{\nabla}T$.
Generally, deviations arise from {\it local} equilibrium either from
changes in the equilibrium propagators $\vec\nabla \hat g^{R,A}_0$
or from the distribution function $\vec\nabla \Phi_0$ with
\begin{equation}
\Phi_0(\vec{R})= 
1 - 2 f \big( \epsilon; T ( \vec{R} ) \big)
 = \tanh\bigg( \frac{\epsilon}{2 k_B T(\vec{R})} \bigg) \,.
\end{equation}

We consider only spatially homogeneous superconducting states
which are in equilibrium, and are
``unitary'', i.e.,\  the equilibrium mean-field order 
parameter satisfies
$\hat{\Delta}(\vec{p}_f)^2 = -|\Delta(\vec{p}_f)|^2\,\hat{1}$,
where $|\Delta|^2$ stands for either the spin scalar product
$\Delta\,\underline{\Delta}$ or the spin vector product
$\vec\Delta\cdot\underline{\vec\Delta}$.

\subsection{Solutions of the linearized quasiclassical transport equations}

The deviations of the propagators from their local equilibrium values,
$\delta\hat{g}^{X}=\hat{g}^{X}-\hat{g}^{X}_0$,
and $\delta\hat{\sigma}^{X}=\hat{\sigma}^{X}-\hat{\sigma}^{X}_0$
with $X\in\{R,A,K\}$, satisfy the following linearized equations
for the retarded and advanced propagators,
\begin{eqnarray}\label{app1}
&&
\left[
  \delta\hat{g}^{R,A} , \, \hat{h}^{R,A}
\right]_\circ = i\partial\hat{g}_0^{R,A} +
\left[
  \hat{g}_0^{R,A} , \, \hat{\sigma}_{ext}+\delta\hat{\sigma}^{R,A}
\right]_\circ \,,
\end{eqnarray}
and for the anomalous propagator,
\begin{eqnarray}\label{app2}
\hat h^R\circ  \delta\hat{g}^{a} \, - \, 
	\delta\hat{g}^{a}\circ\hat h^A &=&
(i\partial\Phi_0)\circ\hat g^A_0
-\hat g^R_0\circ(i\partial\Phi_0)
+\delta\hat{\sigma}^{a}\circ\hat g^A_0 -\hat g^R_0\circ
\delta\hat{\sigma}^{a}
\nonumber  \\
&&
-\left[ \hat{\sigma}_{ext}\, , \,\Phi_0 \right]_\circ \circ\hat g^A_0
-\hat{g}^R_0\circ \left[ \Phi_0\, , \,\hat{\sigma}_{ext}
\right]_\circ \,,
\end{eqnarray}
where $\hat{h}^{R,A}=\epsilon\hat{\tau}_3-\hat{\sigma}_0^{R,A}$, and
$\partial=\hbar\vec{v}_{\!f}\cdot\vec{\nabla}$. We introduced 
Eliashberg's anomalous propagator, $\delta\hat{g}^{a}$, and self-energy,
$\delta\hat{\sigma}^{a}$, defined by
\begin{eqnarray}\label{gK_lin}
&\delta\hat{g}^{K} = \delta\hat{g}^{R} \circ \Phi_0 -\Phi_0 \circ \,
\delta\hat{g}^{A} +\delta\hat{g}^{a}\,,
\\ \label{app6}
&\delta{\hat\sigma}^{K} = \delta\hat{\sigma}^{R} \circ \Phi_0 -\Phi_0
\circ \, \delta\hat{\sigma}^{A} +\delta\hat{\sigma}^{a} \, .
\end{eqnarray}
The transport equations for $\delta\hat{g}^{R,A,a}$ 
are simplified by inserting the local equilibrium propagators
\begin{equation}
\hat{g}^{R,A}_0=
	\frac{ \tilde{\epsilon}^{R,A}\hat{\tau}_3-
	{\hat{\tilde{\Delta}}}^{R,A} } {C^{R,A}} \,,
\quad \mbox{with} \quad
C^{R,A}=-\frac{1}{\pi}\sqrt{| \tilde{\Delta}^{R,A} |^2-
   (\tilde{\epsilon}^{R,A})^2}\,, 
\end{equation}
where $\tilde{\epsilon}^{R,A}$ and ${\tilde{\Delta}}^{R,A}$
determine the equilibrium self-energy,
\begin{equation}
\hat{\sigma}^{R,A}_0=(\epsilon - \tilde{\epsilon}^{R,A})\hat{\tau}_3
+ {\hat{\tilde\Delta}}^{R,A} + D^{R,A}\hat{1}
\,.
\end{equation}

  Finally, the equations for $\delta\hat g^{R,A}$ and $\delta\hat g^a$
are solved by using the normalization conditions
\begin{equation}\label{app10}
\hat{g}_0^{R,A} \circ \delta\hat{g}^{R,A}\,+\,
\delta\hat{g}^{R,A}\circ \hat{g}_0^{R,A}\,=\,0 \enspace, 
\quad\mbox{and}\qquad
\hat{g}_0^{R} \circ \delta\hat{g}^{a}\,+\,
\delta\hat{g}^{a}\circ \hat{g}_0^{A}\,=\,0 \enspace,
\end{equation}
and $\hat{h}^{R,A}$ in order to move the propagators
$\delta\hat g^{R,A}$ and $\delta\hat g^a$  to the right
on the left hand side of (\ref{app1}) and (\ref{app2}).
The solutions are
\begin{eqnarray}\label{gRA_lin}
\delta\hat{g}^{R,A} &=&
\left(
	C^{R,A}_{+} \hat{g}_0^{R,A} + D^{R,A}_{-}
\right)^{-1}\!\! \circ \!
\biggl(
  -i\partial\hat g^{R,A}_0 +
  \left[
    \hat{\sigma}_{ext}+\delta\hat{\sigma}^{{R,A}}\,, \,\hat g^{R,A}_0
  \right]_\circ
\biggr)
\,,
\\ \label{ga_lin}
\delta\hat{g}^{a} &=&
\left(
	C^a_{+} \hat{g}_0^R + D^a_{-}
\right)^{-1}\!\! \circ \!
\biggl(
  (i\partial\Phi_0)\circ\hat g^A_0
  -\hat g^R_0\circ(i\partial\Phi_0)
\nonumber  \\ &&
  +\delta\hat{\sigma}^{a}\circ\hat g^A_0 -\hat g^R_0\circ
  \delta\hat{\sigma}^{a}
  -\left[ \hat{\sigma}_{ext}\, , \,\Phi_0 \right]_\circ \circ\hat
g^A_0
  -\hat g^R_0\circ \left[ \Phi_0\, , \,\hat{\sigma}_{ext}
\right]_\circ
\biggr)\,,
\end{eqnarray}
with the auxiliary functions
\begin{eqnarray}\label{CRA+}
C_+^{R,A}(\vec p_{\!f};\epsilon,\omega) &=&
C^{R,A}(\vec p_{\!f};\epsilon_{+})+C^{R,A}(\vec p_{\!f};\epsilon_{-})
\,,
\\ \label{DRA-}
D_-^{R,A}(\vec p_{\!f};\epsilon,\omega) &=&
D^{R,A}(\vec p_{\!f};\epsilon_{+})-D^{R,A}(\vec p_{\!f};\epsilon_{-})
\,,
\\ \label{Ca+}
C^a_+(\vec p_{\!f};\epsilon,\omega) &=&
C^{R}(\vec p_{\!f};\epsilon_{+})+C^{A}(\vec p_{\!f};\epsilon_{-}) \,,
\\ \label{Da-}
D^a_-(\vec p_{\!f};\epsilon,\omega) &=&
D^{R}(\vec p_{\!f};\epsilon_{+})-D^{A}(\vec p_{\!f};\epsilon_{-}) \,,
\end{eqnarray}
and with $\epsilon_{\pm} = \epsilon\pm \hbar\omega/2$.
Using the normalization conditions (\ref{norm}) we can immediately
invert the matrix
\begin{equation}
\left( C \hat{g}_0^{R,A} + D \right)^{-1} =
-\frac{ C \hat{g}_0^{R,A} - D }{ \pi^2 C^2 + D^2 } \,.
\end{equation}

  After combining the previous results (\ref{gRA_lin}) and (\ref{ga_lin}),
according to (\ref{gK_lin}), we obtain the general expression 
for the linearized Keldysh propagator $\delta\hat{g}^{K}$:
\begin{eqnarray}\label{app14}
\delta\hat{g}^{K} &=&
\left(
	C^{R}_{+} \hat{g}_0^{R} + D^{R}_{-}
\right)^{-1} \circ
\biggl(
  -i\partial\hat g^{R}_0 +
  \left[
    \hat{\sigma}_{ext}+\delta\hat{\sigma}^{{R}}\,, \,\hat g^{R}_0
  \right]_\circ
\biggr) \circ \Phi_0
\nonumber\\ &&
{-}\Phi_0 \circ \left(
	C^{A}_{+} \hat{g}_0^{A} + D^{A}_{-}
\right)^{-1} \circ
\biggl(
  -i\partial\hat g^{A}_0 +
  \left[
    \hat{\sigma}_{ext}+\delta\hat{\sigma}^{{A}}\,, \,\hat g^{A}_0
  \right]_\circ
\biggr)
\nonumber\\ &&
{+}\left(
	C^a_{+} \hat{g}_0^R + D^a_{-}
\right)^{-1} \circ
\biggl(
   (i\partial\Phi_0)\circ\hat g^A_0 -\hat g^R_0\circ(i\partial\Phi_0)
   +\delta\hat{\sigma}^{a}\circ\hat g^A_0 -\hat g^R_0\circ
   \delta\hat{\sigma}^{a} 
\nonumber\\ && \hspace{35mm}
   -\left[ \hat{\sigma}_{ext}\, , \,\Phi_0 \right]_\circ \circ\hat
g^A_0
   - \hat g^R_0\circ \left[ \Phi_0\, , \,\hat{\sigma}_{ext}
\right]_\circ
\biggr)\,.
\end{eqnarray}

  A closer look at (\ref{app14}) shows that changes in the equilibrium
propagators $\partial \hat g^{R,A}_0$ do not contribute to a static
heat current, because the normalization condition (\ref{norm})
gives $2\, {\rm Tr}\, \hat g^{R,A}_0 \partial \hat g^{R,A}_0 =
\partial\, {\rm Tr}\, \big( \hat g^{R,A}_0 \big)^2 = 0.$
  In linear response the physical observables are then determined by 
the deviations from the quasiparticle distribution through
$\delta \hat g^K$.  In general the corrections to the self-energies 
$\delta\hat\sigma$ must be included and require a self-consistent 
calculation.

\end{document}